\documentclass[twocolumn,resetfootnote]{aastex701}

\newcommand{\msun}{M_\odot}
\newcommand{\mpc}{\mathrm{Mpc}}
\newcommand{\mvir}{M_{\rm 200c}}
\newcommand{\rvir}{R_{\rm 200c}}

\newcommand{\mstar}{M_{\rm star}}

\newcommand{\rmagcen}{M_{r,{\rm cen}}}

\newcommand{\disperse}{\textsc{DisPerSE}}
\newcommand{\eg}{e.g.}

\defcitealias{2015ApJ...800..112G}{G15}

\usepackage{multirow}
\usepackage{array}
\usepackage{float}

\begin{document}

\title{The impact of cosmic filaments on the abundance of satellite galaxies}

\newcommand{\NAOC}{National Astronomical Observatories, Chinese Academy of Sciences, Beijing 100101, China}
\newcommand{\UCAS}{School of Astronomy and Space Science, University of Chinese Academy of Sciences, Beijing 100049, China}

\author[0009-0004-6895-6743]{Yuxi Meng}
\affiliation{\NAOC}
\affiliation{\UCAS}
\email[show]{yxmeng@bao.ac.cn}

\author[0000-0002-1665-5138]{Haonan Zheng}
\affiliation{The Kavli Institute for Astronomy and Astrophysics, Peking University, 5 Yiheyuan Road, Haidian District, Beijing 100871, China}
\email[]{}

\author[0000-0001-7075-6098]{Shihong Liao}
\affiliation{\NAOC}
\affiliation{\UCAS}
\email[show]{shliao@nao.cas.cn}

\author[0000-0003-3864-068X]{Lizhi Xie}
\affiliation{Astrophysics Center, Tianjin Normal University, Tianjin 300387, China}
\email[]{}

\author[0000-0002-9788-2577]{Lan Wang}
\affiliation{\NAOC}
\affiliation{\UCAS}
\email[]{}

\author[0009-0001-2951-7383]{Hongxiang Chen}
\affiliation{\NAOC}
\affiliation{\UCAS}
\email[]{}

\author[0009-0006-3885-9728]{Liang Gao}
\affiliation{Institute for Frontiers in Astronomy and Astrophysics, Beijing Normal University, Beijing 102206, China}
\affiliation{School of Physics and Laboratory of Zhongyuan Light, Zhengzhou University, Zhengzhou 450001, China}
\affiliation{\NAOC}
\email[]{}

\author[0000-0003-2858-5090]{Quan Guo}
\affiliation{Shanghai Astronomical Observatory, Chinese Academy of Sciences, Shanghai 200030, China}
\email[]{}

\author[0000-0003-3433-8416]{Yingjie Jing}
\affiliation{\NAOC}
\email[]{}

\author[0000-0002-9937-2351]{Jie Wang}
\affiliation{\NAOC}
\affiliation{\UCAS}
\email[]{}

\author[0000-0003-3279-0134]{Hang Yang}
\affiliation{\NAOC}
\email[]{}

\author[0000-0003-4708-3344]{Guangquan Zeng}
\affiliation{Department of Physics, The Chinese University of Hong Kong, Shatin, N.T., Hong Kong S.A.R., China}
\email[]{}

\begin{abstract}

The impact of cosmic web environments on galaxy properties plays a critical role in understanding galaxy formation. Using the state-of-the-art cosmological simulation IllustrisTNG, we investigate how satellite galaxy abundance differs between filaments and the field, with filaments identified using the DisPerSE algorithm. When filaments are identified using galaxies as tracers, we find that, across all magnitude bins, central galaxies in filaments tend to host more satellite galaxies than their counterparts in the field, in qualitative agreement with observational results from the Sloan Digital Sky Survey. The average ratios between satellite luminosity functions in filaments and the field are $3.49$, $2.61$, and $1.90$ in the central galaxy $r$-band magnitude bins of $M_{r, {\rm cen}} \sim -22$, $-21$, and $-20$, respectively. We show that much of this excess can be attributed to the higher host halo masses of galaxies in filaments. After resampling central galaxies in both environments to match the halo mass distributions within each magnitude bin, the satellite abundance enhancement in filaments is reduced by up to $79 \%$. Additionally, the choice of tracers used to identify filaments introduces a significant bias: when filaments are identified using the dark matter density field, the environmental difference in satellite abundance is reduced by more than $70 \%$; after further resampling in both magnitude and halo mass, the difference is further suppressed by another $\sim 60$--$95 \%$. Our results highlight the importance of halo mass differences and tracer choice biases when interpreting and understanding the impact of environment on satellite galaxy properties.

\end{abstract}



\section{Introduction} \label{sec:intro}

In the standard cosmological constant ($\Lambda$)--cold dark matter ($\Lambda$CDM) model, structures form hierarchically, with smaller structures forming first and merging to create larger ones \citep[see e.g.,][for reviews]{Frenk2012,Zavala2019}. When an isolated dark matter halo falls into the virial radius of a more massive halo, it becomes a subhalo, experiencing dynamical friction \citep{1943ApJ....97..255C} and orbiting the host halo. During this process, subhalos are subject to tidal forces that gradually strip their mass and can eventually disrupt their structure. Over the past decades, high-resolution N-body simulations have shown that several key subhalo properties (e.g., abundance, spatial distribution) strongly depend on the host halo properties \citep[e.g.,][]{Gao2004,Gao2011,Diemand2007,Springel2008,Klypin2011,Jiang2017}. In particular, after scaling to the host halo mass, the subhalo mass function is well described by a universal function \citep[see e.g.,][]{Angulo2009,Boylan-Kolchin2010,Gao2011,Rodriguez-Puebla2016}, indicating that the subhalo abundance is primarily determined by the host halo mass. When baryonic physics is included, galaxies form within halos and subhalos: the central galaxy occupies the center of the host halo, whereas subhalos generally host satellite galaxies. Subhalos and satellite galaxies are closely related to the small-scale challenges of the $\Lambda$CDM model \citep[see][for reviews]{Bullock2017,2022NatAs...6..897S} and represent one of the most important research areas in astrophysics.

At the same time, galaxies and their halos are embedded in the large-scale cosmic web, composed of knots (nodes), filaments, sheets, and voids \citep{Bond1996}. Understanding how these large-scale environments influence galaxy properties is a fundamental question in galaxy formation \citep[e.g.,][]{2007MNRAS.381...41H,2007MNRAS.375..489H,Mo2010}. Among these components, filaments are particularly important: they contain a substantial fraction of the web’s mass \citep[e.g., $\sim 50 \%$, see][]{Cautun2014}, connect the densest regions, and funnel dark matter and baryons into knots, thereby influencing the evolution of the galaxies that reside within them. 

Most previous studies have focused on the impact of filaments on host halos and central galaxies. They find that the spins and shapes of halos (galaxies) residing in filaments correlate with the filament orientation: low-mass halos and galaxies tend to have spins aligned with filaments, while high-mass systems exhibit perpendicular alignments \citep[see e.g.,][]{2007ApJ...655L...5A, 2007MNRAS.381...41H, 2012MNRAS.421L.137L,Dubois2014, Codis2018, Wang2018, GaneshaiahVeena2019, 2025MNRAS.539..487S,2025ApJ...983L...3R,2025MNRAS.539.1692Z,2025JCAP...10..095W}. Moreover, galaxy properties such as color, stellar mass, star formation rate, and gas content are also found to depend on the filament environment \citep[e.g.,][]{2004MNRAS.353..713K, 2005ApJ...629..143B, 2015MNRAS.446.1458M, 2017ApJ...837...16D, 2020MNRAS.498.1839X, 2022ApJ...924..132Z, 2024MNRAS.528.4139H, 2025ApJ...986..193Y,2025arXiv250718614N}. At higher redshifts, massive and dense filaments -- potentially detectable via future Ly$\alpha$ emission observations \citep{Liu2025} -- are suggested to facilitate gas cooling and enhance star formation in dwarf galaxies embedded in them \citep{2019MNRAS.485..464L, 2022MNRAS.514.2488Z}, while at lower redshifts filaments have been found to promote galaxy quenching \citep{2022A&A...657A...9C, 2024A&A...690A.300Z}.

By contrast, the impact of filaments on satellite galaxy systems is less explored, and current results remain debated. Compared to less dense environments (e.g., the field, including sheets and voids), halos of similar mass in denser environments (e.g., filaments) are expected to experience a higher halo merger rate \citep{2009MNRAS.394.1825F}, potentially leading to a greater abundance of satellites. This expectation is supported by observations from \cite{2015ApJ...800..112G} (hereafter \citetalias{2015ApJ...800..112G}), who, using Sloan Digital Sky Survey (SDSS) data, find that the satellite luminosity function (LF) of galaxies in filaments is significantly higher, by a factor of $\sim 2$, than that of galaxies in non-filament environments. A recent study by \citet{2025A&A...700A..65M}, based on dark matter-only cosmological simulations, further supports this picture, confirming that host halos in filaments typically contain more subhalos than those in voids. However, \citet{2015MNRAS.446.1458M}, utilizing the Galaxies-Intergalactic Medium Interaction Calculation (GIMIC) suite of cosmological hydrodynamical simulations \citep{Crain2009}, find that the abundance of satellites exhibits only a weak dependence on the web environment. In addition, they find that the number of subhalos per halo for a given mass shows an even weaker environmental dependence. Thus, the impact of filamentary environments on satellite abundance remains uncertain, and further observational and numerical studies -- particularly with recent hydrodynamical simulations -- are necessary.

In this study, we utilize one of the state-of-the-art cosmological hydrodynamical simulations, IllustrisTNG \citep{2018MNRAS.480.5113M,2018MNRAS.477.1206N,2018MNRAS.475..624N,2019ComAC...6....2N,Pillepich2018,2018MNRAS.473.4077P,2018MNRAS.475..676S}, which allows us to investigate the impact of filaments on satellite abundance with direct insights into the physical origins of environmental effects. The paper is organized as follows. In Section~\ref{sec:methods}, we describe the details of the IllustrisTNG simulations, the filament identification algorithm, and the definition of environment. In Section~\ref{sec:results}, we present the impact of filamentary environment on satellite abundance, including comparisons with observations, the impact of magnitude distributions, and the influence of halo mass distributions. The effects of tracer choice are investigated in Section~\ref{sec:tracer}. Finally, we summarize our findings and conclude in Section~\ref{sec:con}.

\section{Methods} \label{sec:methods}

\subsection{Simulation}

\begin{deluxetable*}{llcccccc}
\tablecaption{
Number of central galaxies (and their associated satellites) in bins of $r$-band absolute magnitude ($\rmagcen$) and in different cosmic web environments. Each bin includes all centrals with $\rmagcen$ within $\pm 0.5$ mag of the bin center. The numbers in parentheses give the total number of satellites (with more than 100 stellar particles and located within the $\rvir$ of their host) associated with the centrals in each bin. The listed stellar mass ranges correspond to the 5th–95th percentile of all central galaxies, irrespective of environment. Note that the centrals in each bin are selected by $\rmagcen$, so their stellar mass ranges slightly overlap. For both tracers, the total number of central galaxies in the combined filament and field environments is the same in each magnitude bin, owing to our environment definition (see Section \ref{subsec:env}).
\label{tab:1}}
\tablehead{
\colhead{\multirow{2}{*}{Bins}} &
\colhead{\multirow{2}{*}{$\mstar$ range}} &
\colhead{\multirow{2}{*}{Total}} &
\colhead{\multirow{2}{*}{Knot}} &
\multicolumn{2}{c}{Galaxy tracers} & 
\multicolumn{2}{c}{Dark matter tracers} \\
\colhead{} & \colhead{} & \colhead{} & \colhead{} &
\colhead{Filament} & \colhead{Field} &
\colhead{Filament} & \colhead{Field}
}
\startdata
$\rmagcen \sim -23 $ & $ 10^{11.5-12.1} \msun $ & 64 (3448)   & 39 (2960) & 23 (469)   & 2 (19)      & 24 (482)   & 1 (6)      \\
$\rmagcen \sim -22 $ & $ 10^{10.8-11.6} \msun $ & 455 (3381)  & 19 (578)  & 235 (2187) & 201 (616)   & 328 (2318) & 108 (485)  \\
$\rmagcen \sim -21 $ & $ 10^{10.3-11.0} \msun $ & 1654 (2354) & 47 (35)   & 379 (974)  & 1228 (1345) & 756 (1345) & 851 (974)  \\
$\rmagcen \sim -20 $ & $ 10^{9.8-10.5} \msun  $ & 2628 (922)  & 104 (23)  & 316 (184)  & 2208 (715)  & 760 (326)  & 1764 (573) \\
\enddata
\end{deluxetable*}

The analysis in this study is based on the IllustrisTNG simulations \citep{2018MNRAS.480.5113M,2018MNRAS.477.1206N,2018MNRAS.475..624N,Pillepich2018,2018MNRAS.475..676S}, a series of cosmological, gravo-magneto-hydrodynamical simulations performed with the \textsc{AREPO} code \citep{2010MNRAS.401..791S}. We use the $z = 0$ snapshot from the TNG100-1 hydrodynamical run, which offers both high resolution and a large volume, enabling robust statistical analyses. This simulation includes $1820^3$ dark matter particles and an equal number of initial gas cells within a periodic box of size 110.7 cMpc (comoving Mpc). The adopted cosmology follows \citet{2016A&A...594A..13P}, with parameters $\Omega_{\Lambda,0}=0.6911$, $\Omega_{\rm m,0}=0.3089$, $\Omega_{\rm b,0}=0.0486$, $\sigma_8=0.8159$, $n_{\rm s}=0.9667$, and $h=0.6774$. The mean baryonic particle/cell mass is $1.4\times10^6~\msun$, and the dark matter particle mass is $7.5\times10^6~\msun$. In terms of spatial resolution, the softening length for collisionless particles (i.e., dark matter and star) at $z=0$ is 740 pc, while the minimum gas softening length is 185 pc.

Halos and subhalos in the simulation are identified using the friends-of-friends \citep[FoF,][]{Davis1985} and \textsc{SUBFIND} \citep{2001MNRAS.328..726S,Dolag2009} algorithms, respectively. We define the host halo mass as $\mvir$, the total mass enclosed within the virial radius $\rvir$, within which the mean density is 200 times the cosmic critical density. The galaxy sample is based on the \textsc{SUBFIND} subhalo catalogue in the TNG100-1 run. We exclude subhalos of non-cosmological origin (e.g., fragments or clumps formed through baryonic processes within existing galaxies, which are marked with SubhaloFlag=0 in the SUBFIND subhalo catalog; see section 5.2 in \citealt{2019ComAC...6....2N}) as well as low-mass subhalos containing fewer than 100 star particles, which corresponds to satellite galaxies with stellar masses of $\mstar \sim 10^8~\msun$ or $r$-band magnitudes $M_{r} \sim -14.6$ (with a typical scatter of $\Delta M_{r} = 0.6$). In this study, the stellar mass and $r$-band magnitude of each SUBFIND structure are computed using star particles within a 3D radial aperture of 30 physical kpc, consistent with the SDSS Petrosian aperture. To compute the satellite abundance of each FoF halo, we construct corresponding satellite samples. The central galaxy of each FoF halo is defined as the most bound \textsc{SUBFIND} subhalo in the catalogue, while satellite galaxies are defined as the remaining subhalos that meet the above criteria and reside within the virial radius of their host halo.

For a more direct comparison with the observational results in \citetalias{2015ApJ...800..112G}, we consider the SDSS $r$-band absolute magnitude of galaxies, $M_r$, in this study. We use $M_r$ data from \citet{2018MNRAS.475..624N}, which are based on a model that accounts for internal dust attenuation arising from the distribution of neutral gas and metals within galaxies (specifically, `Model C' in \citealt{2018MNRAS.475..624N}).
\footnote{We have tested and confirmed that our main conclusions are not sensitive to the dust model and remain consistent when using dust-free $M_r$.}
We include all central galaxies with magnitudes in the range $-23.5 \leq \rmagcen \leq -19.5$, and divide them into four magnitude bins. The bright-end limit of $\rmagcen = -23.5$ is motivated by \citetalias{2015ApJ...800..112G}, while the faint-end limit of $\rmagcen = -19.5$ is chosen to maximize the sample size within the range of reliable numerical resolution. 
The third column of Table~\ref{tab:1} lists the total number of central galaxies in each bin (with the corresponding satellite counts given in parentheses), while the second column provides the typical stellar mass range of centrals.

\subsection{Filament identification} \label{subsec:filament}

\begin{figure*}[]
\plotone{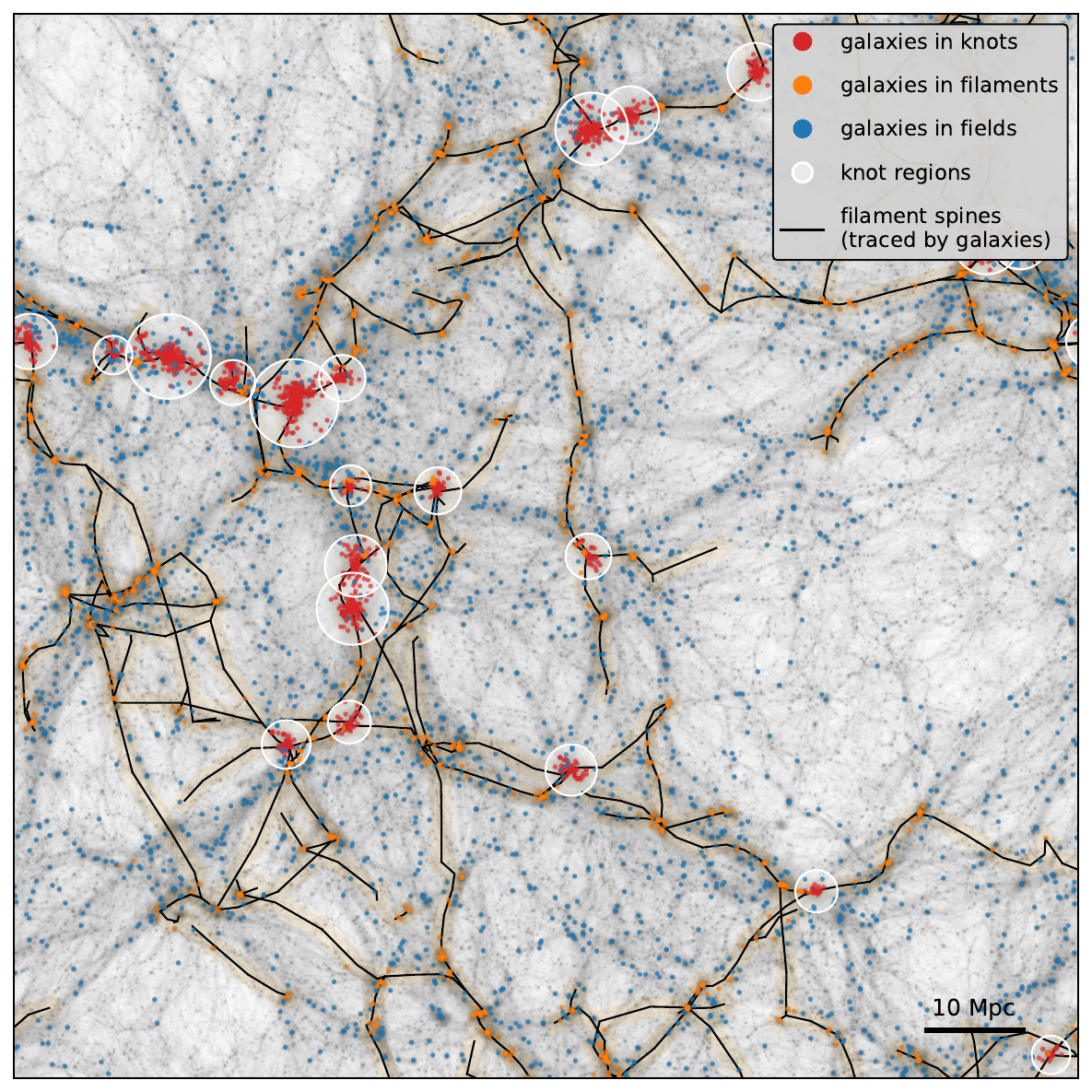}
\caption{Filaments in a $30$ Mpc-thick slice of the TNG100-1 simulation, identified by the \disperse~algorithm using galaxies with stellar masses $\mstar \geq 10^{9}~\msun$ as tracers. The grayscale background shows the dark matter density field. White circles mark the $3\times\rvir$ vicinities of galaxy clusters with $\mvir \geq 10^{13.5}~\msun$, which are defined as the knot environment. Red dots indicate the centers of galaxies residing in knots, which are excluded from our analysis. Black solid line segments show the filament spines defined by \disperse. Orange dots mark the centers of galaxies located within cylinders of radius of $R_{\rm filament} = 1$ Mpc around the filament spines (indicated by the orange shaded regions), and defined as filament galaxies. The remaining galaxies, shown as blue dots, are field galaxies. Visually, despite some minor discrepancies, the filament spines overall capture the massive filamentary structures in the underlying dark matter density field.
\label{fig:M1}}
\end{figure*}

In this study, we identify filaments using the \disperse~algorithm \citep{2011MNRAS.414..350S,2011MNRAS.414..384S}, a widely used scale-free filament finder applicable to both simulation and observational data \citep[see e.g.,][for comparisons of different filament identification methods]{2018MNRAS.473.1195L,2020MNRAS.493.1936R}. Below, we briefly describe how \disperse~works and refer readers to the original paper for theoretical details \citep{2011MNRAS.414..350S}. 

\disperse~first estimates the underlying density field from the distribution of input discrete tracers (e.g., galaxies or dark matter particles; note that galaxies are not mass-weighted). It then identifies critical points (maxima, saddle points, and minima) and traces ridge lines connecting maxima and saddle points to define filament spines. Each filament is assigned a persistence ratio, defined as the density ratio between the connected maximum and saddle point. By applying a persistence threshold (number of $\sigma$s, in analog to Gaussian random field), \disperse~filters out spurious structures and retains only robust filaments. Higher thresholds retain only the most prominent filaments, while lower thresholds include more, potentially weaker, structures. In this context, persistence effectively serves as a signal-to-noise measure for detected filaments. Optional smoothing can be applied to both the estimated density field and the filament spines: the former reduces small-scale shot noise in the density field by averaging densities over neighboring tracers, while the latter suppresses nonphysical sharp edges in the filament geometry by averaging the positions of spine points \citep[see e.g.,][]{2020A&A...642A..19M, 2020A&A...634A..30M}. 

Following observations, we use galaxies as tracers for filament identification. Similar to \citet{2024A&A...684A..63G}, we select galaxies according to observational limits, balancing the identification of real structures with the suppression of spurious ones. Specifically, we trace filaments using galaxies (including both central and satellite galaxies) with stellar masses $\mstar \geq 10^9~\msun$, apply smoothing on both the estimated density field and the identified filaments (each smoothed once), and select filaments with a persistence ratio above the threshold of $2\sigma$. Note that we have tested a range of persistence ratio thresholds (from $1\sigma$ to $6\sigma$, see Appendix~\ref{app:filament_tests} for details), with and without smoothing procedures applied, and confirmed that the conclusions presented in the following sections are not sensitive to these choices.

The identified filament spines in a slice of the simulation box are shown as black solid lines in Figure~\ref{fig:M1}. Overall, the filament spines capture the massive filamentary structures of the underlying dark matter density field (shown as the grayscale background), despite minor mismatches in smaller features. These discrepancies arise primarily from the use of a 2D projection and from galaxies being a sparse tracer of the density field. The impact of tracers will be examined in Section~\ref{sec:tracer}.

\subsection{Definition of environment} \label{subsec:env}

After identifying the filamentary spines, we further classify central galaxies into `filament' and `field' environments. To do this, we first exclude the galaxies residing in clusters (i.e., the `knot' environment), as these environments have distinct effects on galaxy properties compared to filaments and fields. Specifically, central galaxies located within $3 \rvir$ of clusters with $\mvir \geq 10^{13.5}~\msun$ are classified as `knot' environment galaxies (i.e., galaxies marked by red dots in Figure~\ref{fig:M1}) and are excluded from our galaxy samples. The numbers of `knot' galaxies in different $\rmagcen$ bins are listed in the fourth column of Table~\ref{tab:1}.

To assign central galaxies to filaments, we adopt the conclusion from \citet{2024MNRAS.532.4604W} that the typical filament boundary lies at $\sim 1~\mpc$, and identify a galaxy as part of a filament if its center lies within a cylinder of radius $R_{\rm filament} = 1~\mpc$ around the filament spine \citep[see also the discussions in][]{2025ApJ...989..187Y}. Central galaxies not assigned to either filaments or knots are classified as field galaxies. In this work, the environment is defined for each central galaxy and its host halo. Since our analysis concerns the satellites associated with their corresponding central galaxies, satellites are not classified separately. Instead, they are assigned the environment of the centrals. The distribution of galaxies used as tracers, along with their assigned cosmic web environments (orange for filament and blue for field), is shown in Figure~\ref{fig:M1}. The numbers of central galaxies in each magnitude bin for the filament and field environments are given in the fifth and sixth columns of Table~\ref{tab:1}, respectively.

We have also tested alternative parameter choices for the definitions of knots and filaments, such as varying the halo mass threshold representing knot regions ($\mvir \geq 10^{13}$ or $10^{14}\msun$), the knot region radius ($2$--$5\times\rvir$), and the filament radius ($R_{\rm filament} = 0.5$--$3.0~\mpc$). These parallel analyses yield qualitatively similar results. Therefore, we present only the results corresponding to our fiducial parameter set in the following sections.

\section{Satellite luminosity functions} \label{sec:results}

\subsection{Environmental dependence} \label{subsec:obs}

\begin{figure*}[]
\centering\includegraphics[width=\textwidth]{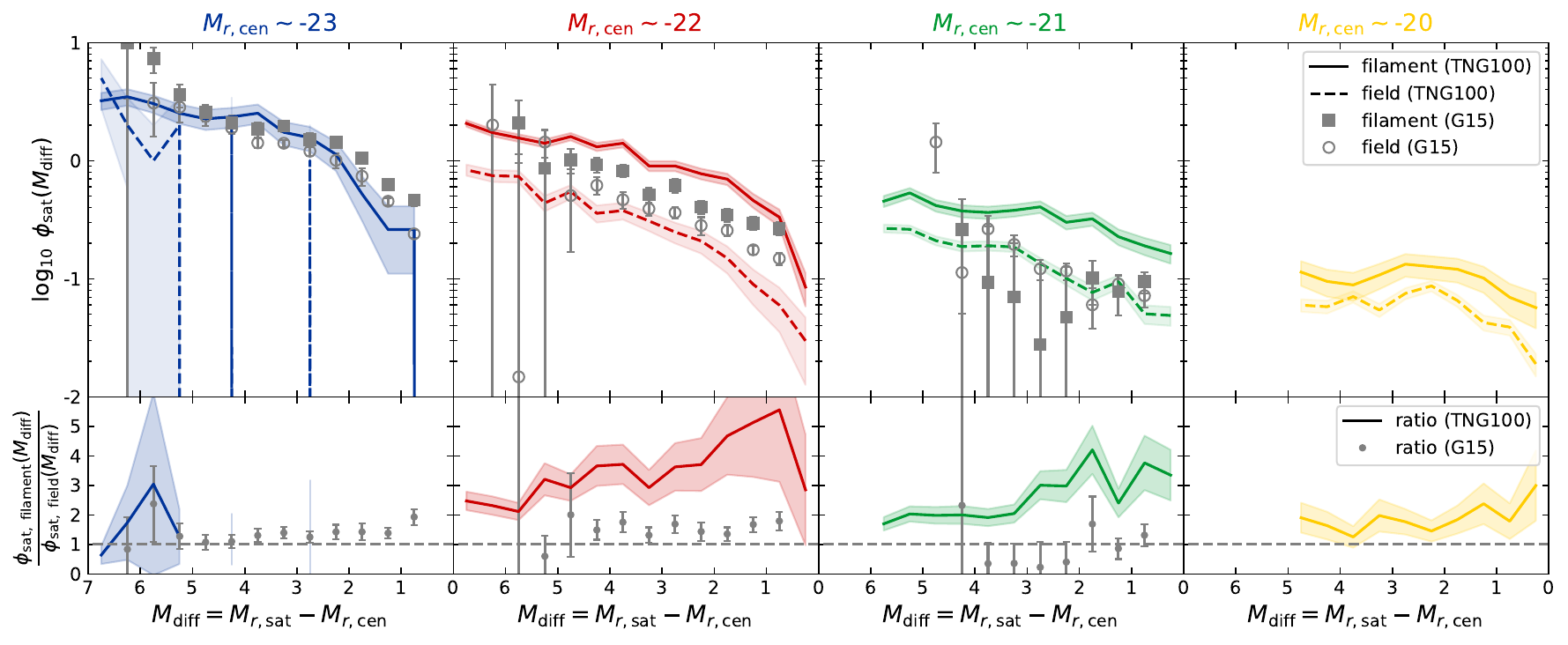}
\caption{Environmental dependence of satellite LFs. From left to right, each column shows the results for central galaxies in the bins of $\rmagcen \sim -23$ (blue), $-22$ (red), $-21$ (green), and $-20$ (yellow). In the upper panel of each column, solid and dashed lines represent simulated satellite LFs for centrals in filaments and fields, respectively. Shaded regions indicate Poisson errors computed using Eq. (\ref{eq:poisson_error}). Filled squares and open circles show observational results from \citetalias{2015ApJ...800..112G}. Bottom panels display the ratio of the filament to field satellite LFs, with errors propagated accordingly. The satellite LF in the $\rmagcen \sim -23$ bin for field galaxies is sparsely sampled due to the limited number of objects, and is therefore excluded from the following analysis. For the three remaining bins, central galaxies in filaments tend to host more satellites than their counterparts in the field, in qualitative agreement with the observational results of \citetalias{2015ApJ...800..112G}.
\label{fig:R1}}
\end{figure*}

The satellite LF quantifies the abundance of satellite galaxies as a function of their luminosity, normalized to that of their central galaxy. In the top row of Figure~\ref{fig:R1}, we show the average satellite LFs for central galaxies in filaments (solid lines) and fields (dashed lines), separated into different $\rmagcen$ bins. 
The horizontal axis represents the magnitude difference between a satellite and its central galaxy, defined as $M_{\rm diff} = M_{r, {\rm sat}} - \rmagcen$, which is related to the luminosity ratio between the satellite and its central galaxy. Note that $M_{\rm diff}$ is typically a positive value, as the central galaxy is by definition the brightest galaxy in a group. For satellites of a given central galaxy, a smaller (larger) $M_{\rm diff}$ corresponds to a brighter (fainter) satellite. The vertical axis shows the average number of satellites per central galaxy per unit $M_{\rm diff}$ bin, calculated as
\begin{equation}
    \phi_{\rm sat} (M_{\rm diff}) = \frac{\sum_{i = 1}^{N_{\rm cen}} N_{{\rm sat}, i}(M_{\rm diff})}{N_{\rm cen} \Delta M_{\rm diff}}, 
\end{equation}
where $N_{{\rm sat}, i}(M_{\rm diff})$ is the number of satellites associated with the $i$-th central galaxy whose magnitude differences fall within the bin $[M_{\rm diff} - \Delta M_{\rm diff}/2, M_{\rm diff} + \Delta M_{\rm diff}/2)$, $\Delta M_{\rm diff}$ denotes the bin width of the magnitude difference, and $N_{\rm cen}$ is the total number of central galaxies considered. We also estimate the uncertainty in the satellite LF, shown as shaded regions, based on Poisson errors, i.e.,
\begin{equation}\label{eq:poisson_error}
    \delta \phi_{\rm sat} (M_{\rm diff}) = \frac{\sqrt{\sum_{i = 1}^{N_{\rm cen}} N_{{\rm sat}, i}(M_{\rm diff})}}{N_{\rm cen} \Delta M_{\rm diff}}.
\end{equation}

To compare with observations, we overplot the observed satellite LFs from \citetalias{2015ApJ...800..112G} in gray, using filled squares for filament galaxies and open circles for field galaxies from the SDSS. Note that \citetalias{2015ApJ...800..112G} provide observational results only for the three brightest $\rmagcen$ bins, and the $\rmagcen \sim -21$ bin is affected by significant noise due to the small sample size.

As expected, moving from brighter to fainter central galaxies (from left to right panels), the abundance of satellites in the TNG100 simulations generally decreases in both filament and field environments. This trend is consistent with the observational results from \citetalias{2015ApJ...800..112G}.

In the brightest central galaxy bin ($\rmagcen \sim -23$), there is only two central galaxy in our field sample. As a result, the corresponding satellite LF is sparsely sampled. Given the poor sampling of field central galaxies in this bin, we focus on the three fainter central galaxy bins in the following analysis to compare environmental differences more robustly.

For all three remaining bins in the simulation, we find that central galaxies in filaments tend to host more satellites than their counterparts in the field, in qualitative agreement with the trend in observational results of \citetalias{2015ApJ...800..112G}. This also echoes the similar environmental dependence of subhalo mass functions reported by \citet{2025A&A...700A..65M}. In each $\rmagcen$ bin, the environmental effect is more pronounced for brighter satellites (i.e., those with smaller $M_{\rm diff}$). When comparing across the $\rmagcen$ bins, the difference between filament and field environments decreases toward fainter central galaxies. Specifically, in the $\rmagcen \sim -22$ bin, the satellite abundance in filament is $\sim 2$--$6$ times higher than in the field, with an average ratio of $3.49$. The ratio decreases to $\sim 2$--$4$ (average $2.61$) in the $\rmagcen \sim -21$ bin and $\sim 1$--$3$ (average $1.90$) in the $\rmagcen \sim -20$ bin. These average ratios are summarized in the third column of Table~\ref{tab:2}.

\begin{figure*}[]
\centering\includegraphics[width=\textwidth]{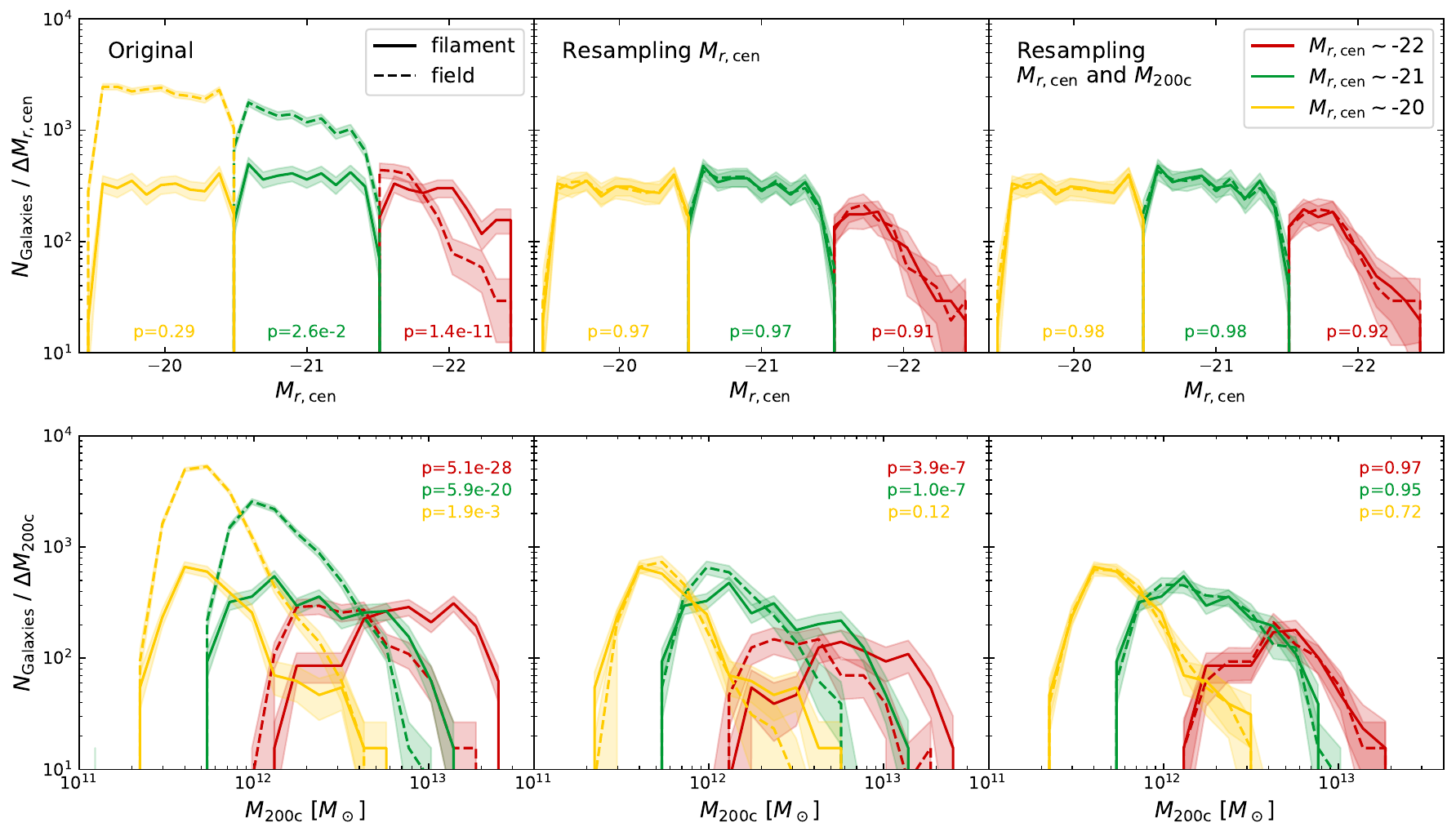}
\caption{{\it Top}: Distributions of central galaxy magnitudes ($\rmagcen$). From left to right, we show the distributions of the original samples, the samples after resampling $\rmagcen$, and the samples after resampling both $\rmagcen$ and $\mvir$. In each panel, as in other figures in this study, the $\rmagcen \sim -22$, $-21$, and $-20$ bins are plotted in red, green, and yellow, respectively. Central galaxies in filaments and the field are distinguished using solid and dashed lines, respectively. For each $\rmagcen$ bin, the Kolmogorov–Smirnov (KS) test $p$-value is shown in the panel, quantifying the similarity between the filament and field $\rmagcen$ distributions. {\it Bottom:} Same as the top panels, but showing the distributions of the host halo virial masses ($\mvir$) for central galaxies. After resampling, the filament and field samples exhibit similar distributions in both $\rmagcen$ and $\mvir$. See the main text for details of the resampling procedure.
\label{fig:R2}}
\end{figure*}

Overall, the environmental difference is quantitatively more pronounced in simulations compared to the \citetalias{2015ApJ...800..112G} observations. This discrepancy could originate from several factors: (i) In simulations, central and satellite galaxies are very well-defined, while in observations, their identification can be affected by the uncertainties in galaxy redshifts and projection effects. Specifically, in \citetalias{2015ApJ...800..112G}, central galaxies are defined to have a characteristic radius, $R_{\rm inner}$, which is comparable to the virial radius of galaxies in the corresponding $\rmagcen$ bin. Galaxies are selected as centrals if, within a projected distance of $2R_{\rm inner}$, all other galaxies are either more than half a magnitude fainter or have a spectroscopic redshift difference larger than 0.002 (or a photometric redshift difference larger than 2.5 times the photometric error). Satellite galaxies are more difficult to define, as they are fainter; thus, a statistical background subtraction technique is adopted to estimate their abundance. We refer interested readers to \citetalias{2015ApJ...800..112G} for more details. The key message is that the definition of central and satellite galaxies in observations is more challenging, which could potentially affect the final results. (ii) The use of different filament identification methods could also contribute to the differences. \citetalias{2015ApJ...800..112G} identify filament spines using the Bisous method \citep{2014MNRAS.438.3465T}, an object point process model that detects and characterizes filamentary structures in the cosmic web by using a network of small, interacting cylindrical segments to trace filaments in galaxy distributions. Here, we use the \disperse~finder, which is more widely used, open-source, and enables us to study the effects of different tracers. The differences in methods and parameter choices could also lead to quantitative discrepancies in the results. 

In this work, we do not aim to fully reproduce the observations by strictly following the same procedures. The intriguing result is that the simulations exhibit a qualitatively similar environmental trend to that observed. This suggests that the trend is not tied to a particular choice of galaxy classification or filament identification, but instead emerges consistently, even though the methods employed here differ from those in \citetalias{2015ApJ...800..112G}. In the following subsections, we further examine the possible origin of this environmental difference.

\begin{figure*}[]
\centering\includegraphics[width=\textwidth]{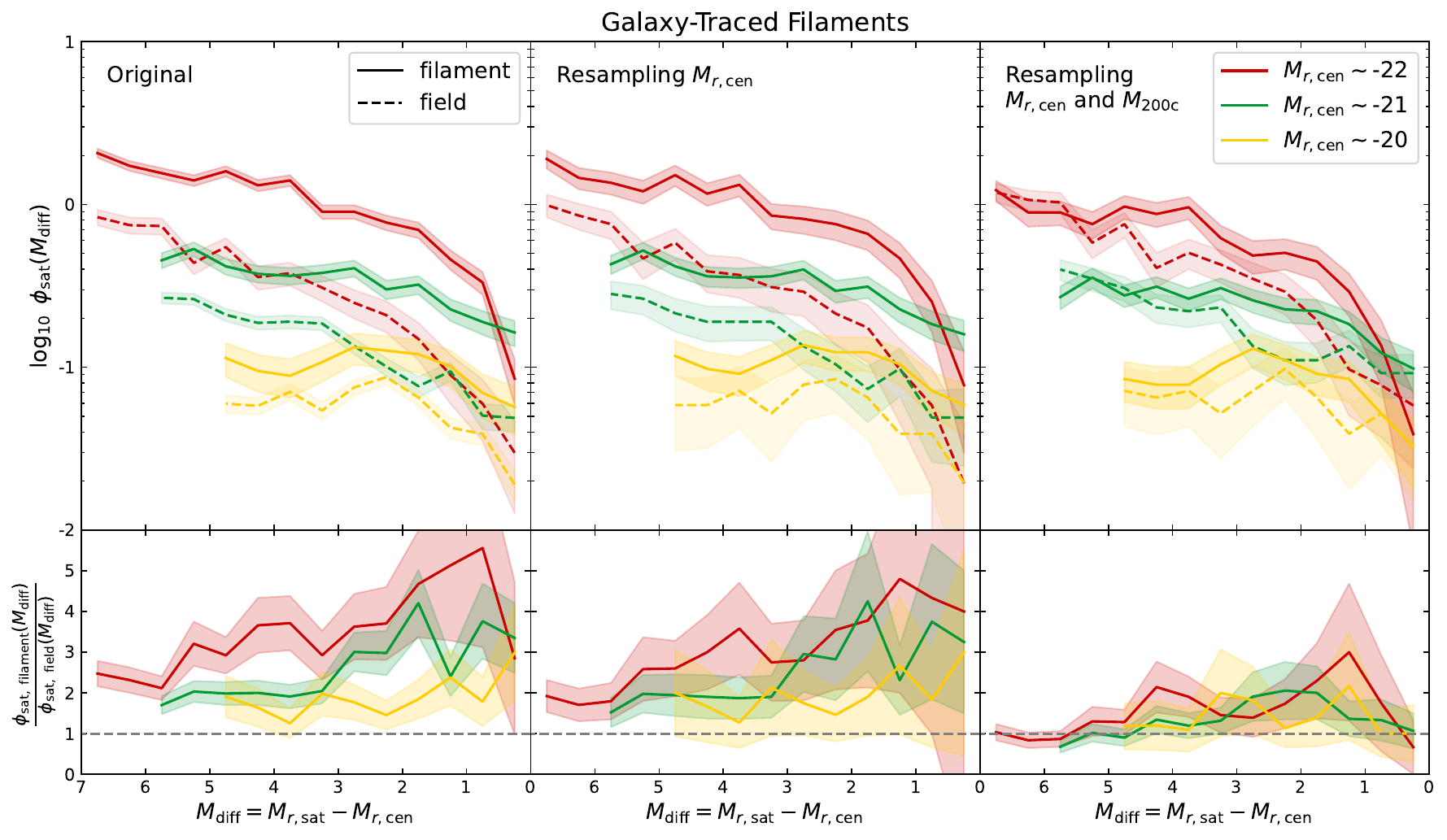}
\caption{Impact of magnitude and halo mass distributions on the environmental differences in satellite LFs. {\it Left}: The upper panel summarizes the satellite LFs for the $\rmagcen \sim -22$ (red), $-21$ (green), and $-20$ (yellow) bins from Figure~\ref{fig:R1}. Results for filament and field central galaxies are shown with solid and dashed lines, respectively. The lower panel shows the corresponding ratios between filament and field satellite LFs. {\it Middle}: Same as the left panels, but for galaxy samples resampled to have similar $\rmagcen$ distributions. {\it Right}: Same as the middle panels, but for galaxy samples resampled to match both $\rmagcen$ and $\mvir$ distributions. After resampling $\rmagcen$, the average ratios decrease by less than $17 \%$. However, after additionally resampling $\mvir$, the average ratios drop by $78 \%$, $79 \%$, and $53 \%$ for the $\rmagcen \sim -22$, $-21$, and $-20$ bins, respectively, relative to the original unresampled samples. This indicates that differences in halo mass distributions between filament and field central galaxies contribute significantly to the environmental differences in satellite abundance, especially in the brighter central galaxy bins.
\label{fig:R3}}
\end{figure*}

\begin{deluxetable*}{ccccc}
\tablecaption{Average satellite abundance ratios between filament and field central galaxies, $\phi_{\rm sat, filament} / \phi_{\rm sat, field}$. 
The values are calculated by averaging the ratios over different $M_{\rm diff}$ bins shown in Figures~\ref{fig:R3} and \ref{fig:R3dm}. For the resampled cases, the reductions of the ratios compared to the original cases are given in brackets, where the fractional reduction is computed as $1 - \left( \frac{\phi_{\rm sat, filament}}{\phi_{\rm sat, field}} \big|_{\rm resampled} - 1 \right)\big/ \left( \frac{\phi_{\rm sat, filament}}{\phi_{\rm sat, field}} \big|_{\rm original} - 1 \right)$. These results highlight that both halo mass and tracer choice significantly affect the measured environmental differences.\label{tab:2}}
\tablehead{
\colhead{\multirow{2}{*}{Filament tracers}} &
\colhead{\multirow{2}{*}{Bins}} &
\multicolumn{3}{c}{$\phi_{\rm sat, filament} / \phi_{\rm sat, field}$} \\
\colhead{} & \colhead{} &
\colhead{Original} &
\colhead{Resampling $\rmagcen$} &
\colhead{Resampling $\rmagcen$ and $\mvir$}
}
\startdata
\multirow{3}{*}{Galaxies}
& $\rmagcen \sim -22$ & $3.49$ & $3.07~(17\%\downarrow)$ & $1.55~(78\%\downarrow)$ \\
& $\rmagcen \sim -21$ & $2.61$ & $2.54~(5\%\downarrow)$  & $1.34~(79\%\downarrow)$ \\
& $\rmagcen \sim -20$ & $1.90$ & $1.92~(2\%\uparrow)$ & $1.43~(53\%\downarrow)$ \\
\tableline
\multirow{3}{*}{Dark matter density}
& $\rmagcen \sim -22$ & $1.67$ & $1.34~(50\%\downarrow)$ & $1.03~(95\%\downarrow)$ \\
& $\rmagcen \sim -21$ & $1.48$ & $1.38~(20\%\downarrow)$ & $1.02~(96\%\downarrow)$ \\
& $\rmagcen \sim -20$ & $1.26$ & $1.21~(17\%\downarrow)$ & $1.10~(60\%\downarrow)$ \\
\enddata
\end{deluxetable*}

\subsection{Impact of magnitude distributions} \label{subsec:magnitude}

In the previous analysis, we adopted a bin width of $\Delta \rmagcen = 1$ for central galaxies. A natural question is whether the observed environmental difference arises from the different distributions of $\rmagcen$ within each bin between filament and field central galaxies. For example, if central galaxies in filaments have a higher fraction of brighter ones, they would naturally host more satellites.

To examine this, we first plot the histograms of central galaxy magnitudes in the upper-left panel of Figure~\ref{fig:R2} for the $\rmagcen \sim -22$, $-21$, and $-20$ bins. Solid and dashed lines show the results for central galaxies in filaments and the field, respectively. Within each $\rmagcen$ bin, central galaxies in fields tend to have a higher probability of being fainter, with the trend becoming more pronounced in the brighter $\rmagcen$ bins. In contrast, central galaxies in filaments exhibit a relatively flat $\rmagcen$ distribution in the $-20$ and $-21$ bins, while in the $-22$ bin, they have a slightly higher probability of being fainter, resembling their field counterparts. Overall, central galaxies in filaments do show a higher fraction of brighter members compared to those in the field, which can partially explain their higher satellite abundance.

To further quantify the impact of magnitude distribution on the environmental difference observed in Section~\ref{subsec:obs}, we resample the central galaxies based on their $\rmagcen$, ensuring that the resampled samples have similar distributions in both environments. Specifically, for each pair of filament and field histograms, we divide them into 20 narrower bins.\footnote{We have tested other choices for the number of narrow bins (e.g., 10, 40) and confirmed that the results remain consistent. A similar test is also performed for the resampling of $\mvir$ (Section~\ref{subsec:halo mass}).} In each narrow bin, central galaxies from the environment with a higher number count are randomly resampled to match the number of galaxies in the other environment. The $\rmagcen$ distributions of the resampled central galaxy samples are shown in the upper-middle panel of Figure~\ref{fig:R2}. We can clearly see that, after resampling, the filament and field histograms are closely matched, as further confirmed by the Kolmogorov–Smirnov (KS) tests shown in Figure~\ref{fig:R2} (i.e., a higher $p$-value indicating that the two distributions are statistically indistinguishable). We then re-examine the environmental differences in satellite LFs using these resampled central galaxies.

The satellite LFs computed from resampled central galaxies across different magnitude bins are plotted in the middle column of Figure~\ref{fig:R3}. Compared to the original results shown in the left column, the resampled LFs exhibit larger Poisson errors due to reduced number counts. However, the ratio panels reveal that the environmental differences remain largely unchanged, with average ratios of $3.07$, $2.54$, and $1.92$ for the $\rmagcen \sim -22$, $-21$, and $-20$ bins, respectively (see the fourth column in Table~\ref{tab:2}). Compared to the original ratios, the resampled values decrease by less than $17 \%$. This suggests that the difference in magnitude distributions accounts for only a small part of the observed environmental effect, and that the environmental difference reported in Section~\ref{subsec:obs} likely arises from other factors.

\subsection{Impact of halo mass distributions} \label{subsec:halo mass}

Another factor that can impact the abundance of satellite galaxies is the host halo mass, $\mvir$ \citep[e.g.,][]{Angulo2009,Boylan-Kolchin2010,Gao2011,Rodriguez-Puebla2016}. For example, if central galaxies in filaments reside more frequently in high-mass halos, they would naturally host more satellites. Unlike observations, hydrodynamical simulations provide precise measurements of host halo virial masses, allowing us to directly examine the impact of halo mass distributions.

In the bottom-left panel of Figure~\ref{fig:R2}, we show histograms of the host halo masses of central galaxies across different environments and magnitude bins. Compared to the field, central galaxies in filaments are indeed more likely to reside in higher-mass halos, and this trend is especially pronounced in the brighter magnitude bins. 

The bottom-middle panel presents the halo mass distributions after resampling the central galaxies to match their $\rmagcen$ distributions. After this magnitude matching, the halo mass distributions in the two environments become somewhat closer, though notable differences remain. Specifically, in the $\rmagcen \sim -20$ and $-21$ bins, filament galaxies tend to reside in more massive halos. In the $-22$ bin, the difference becomes more significant -- filament galaxies are associated with more high-mass halos and fewer low-mass halos than their field counterparts. This indicates that even after matching the magnitude distributions, the halo mass distributions between filaments and the field remain substantially different, which could contribute to the environmental differences observed in the satellite LFs.

To quantify the impact of halo mass distributions, we follow the procedure described in Section~\ref{subsec:magnitude}. Starting from the resampled central galaxy samples with matched $\rmagcen$ distributions, we further divide the halo mass distributions (shown in the bottom-middle panel of Figure~\ref{fig:R2}) into 20 narrower bins. Within each narrow bin, we randomly resample central galaxies in the two environments to ensure equal numbers. This procedure yields central galaxy samples with very similar distributions in both $\rmagcen$ and $\mvir$, as shown in the right column of Figure~\ref{fig:R2}.

With this newly resampled central galaxy samples, we again compute the satellite LFs and quantify the environmental differences. The results are shown in the right column of Figure~\ref{fig:R3}. We find that the environmental differences are now substantially reduced. The average satellite abundance ratios decrease to $1.55$, $1.34$, and $1.43$ for the $M_{r ,{\rm cen}} \sim -22$, $-21$, and $-20$ bins, respectively (see the fifth column of Table~\ref{tab:2}). Compared to the original samples, these values represent reductions of $78 \%$, $79 \%$, and $53 \%$ , respectively. This suggests that the differences in halo mass distributions between filament and field central galaxies make a significant contribution to the environmental differences in satellite abundance observed in the original samples, especially in the brighter central galaxy bins.

\subsection{Discussions}

As shown in the previous subsections, the environmental differences in satellite abundances are partly driven by variations in both central galaxy magnitude and host halo mass distributions. Specifically, after resampling to match the $\rmagcen$ distributions in the two environments, the differences in their satellite LFs are reduced by up to $17 \%$. When the central galaxies are further resampled to match the $\mvir$ distributions, the environmental differences in satellite LFs decrease by up to $79 \%$. This has key implications for interpreting observational results:

Our results suggest that difference in host halo mass distributions (i.e., filaments could host more massive halos compared to the field) can account for up to half of the observed environmental effects. However, measuring host halo masses precisely in observations is challenging.

The remaining environmental differences can be explained by two primary factors: the environmental dependence of halo merger rates \citep{2009MNRAS.394.1825F} and the choice of tracers in filament identification. (i) Satellite galaxies originate from progenitor halos that have merged into a larger host and survived tidal disruption, so their abundance reflects the host halo’s merger history. Because merger rates correlate with large-scale density, and filaments are overdense compared to the field, halos in filaments are expected to experience more frequent mergers, naturally leading to a higher abundance of satellites. (ii) As we will show in the following section, the choice of tracer used to identify filaments also plays a significant role in the measured environmental differences in satellite abundance. In particular, using galaxies as tracers may enhance the environmental dependence of satellite LFs.

\section{Tracer Effects}\label{sec:tracer}

\begin{figure*}[]
\plotone{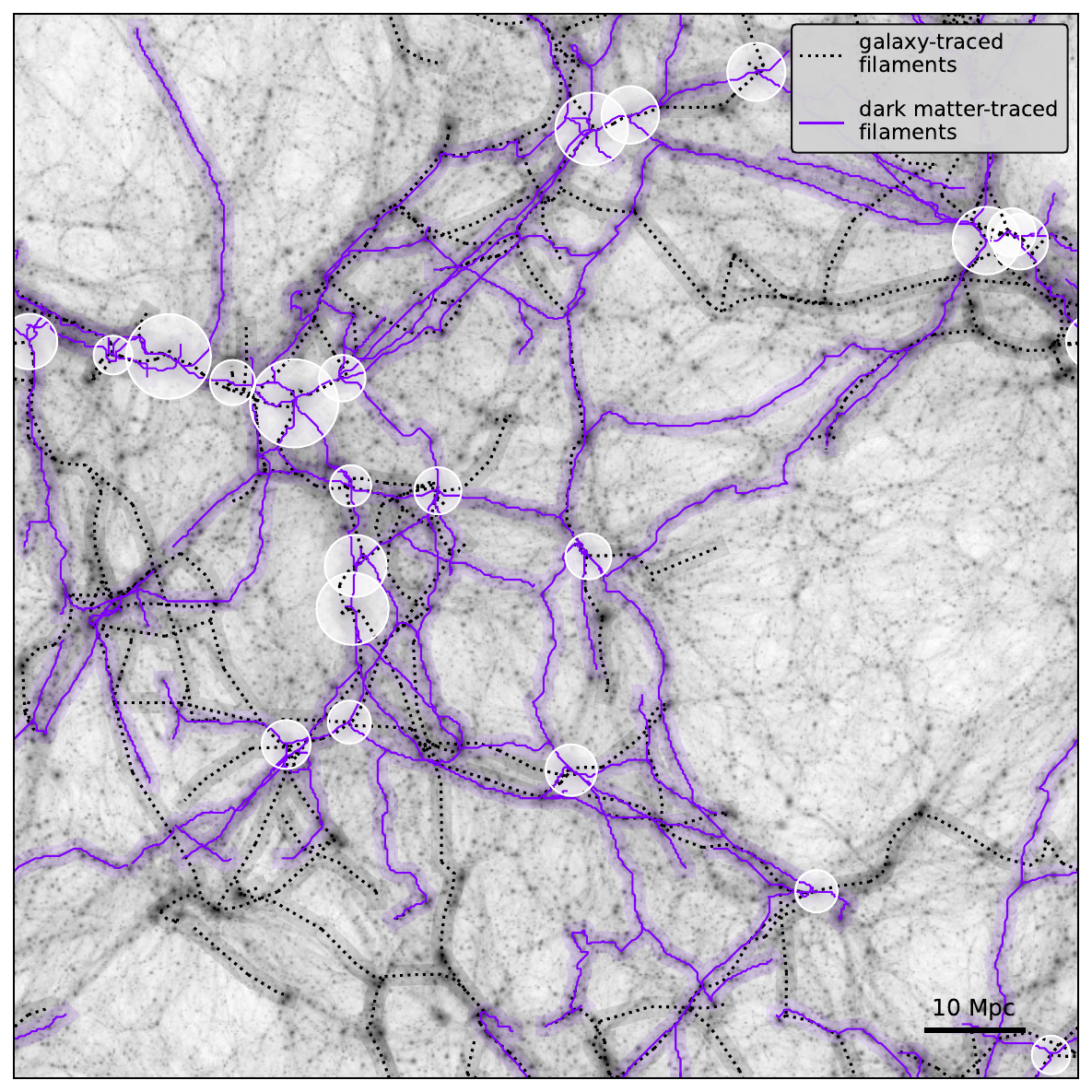}
\caption{Similar to Figure~\ref{fig:M1}, but for filaments identified using dark matter density as tracers. The spines of dark matter-traced filaments are shown as purple solid lines, while the galaxy-traced filament spines from Figure~\ref{fig:M1} are overlaid using black dotted lines for comparison. Filaments identified by these two tracers are qualitatively similar, both following the underlying filamentary matter distribution, but they differ in detailed structures. We investigate these differences and their impact in Section~\ref{sec:tracer}.
\label{fig:M1dm}}
\end{figure*}

\begin{figure*}[]
\centering\includegraphics[width=\textwidth]{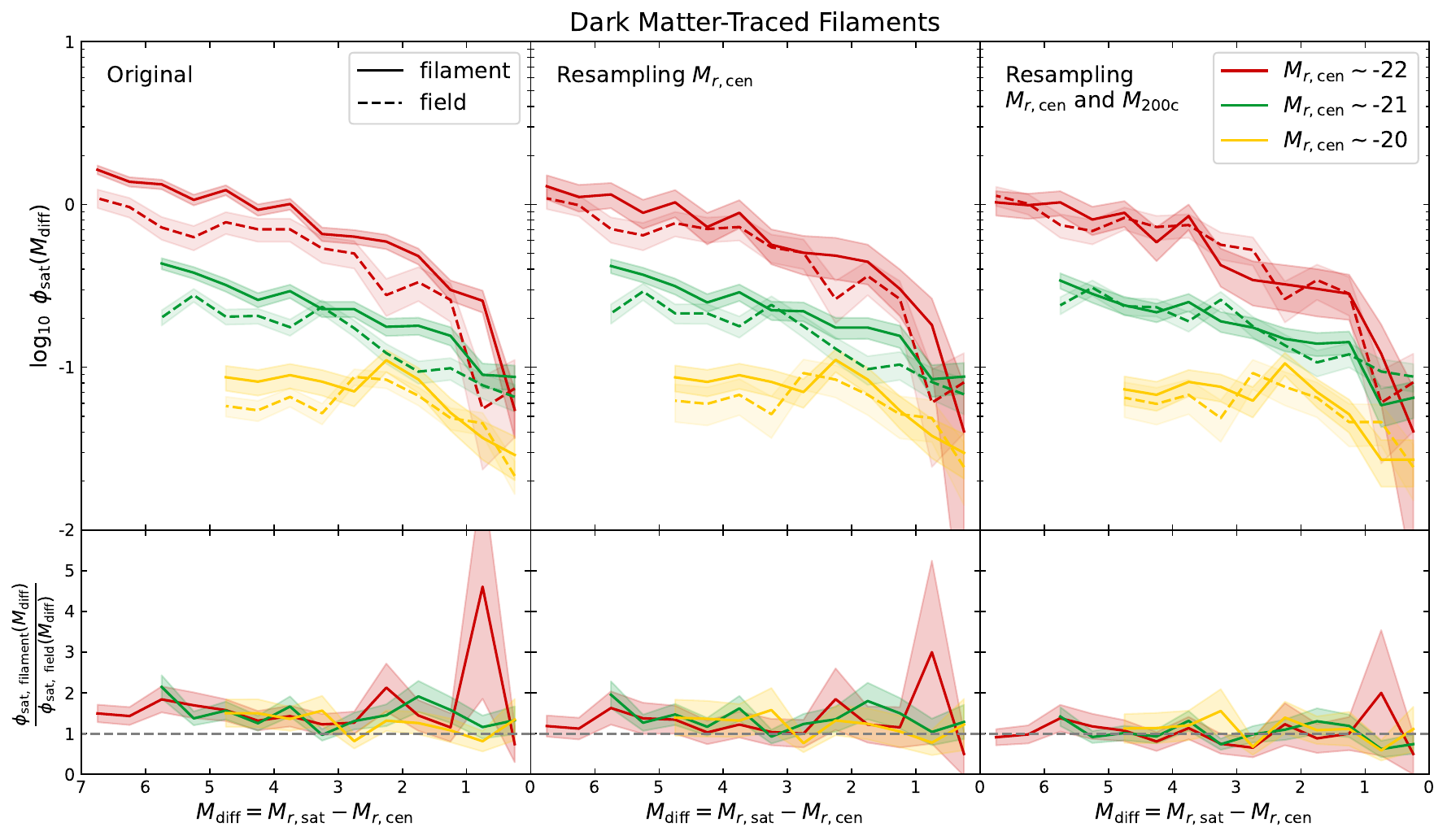}
\caption{Same as Figure~\ref{fig:R3}, but for the parallel analysis where filaments are identified using dark matter density as the tracer. Compared to Figure~\ref{fig:R3}, where filaments are identified using galaxies, the environmental differences are significantly smaller here, suggesting that the choice of filament tracer also influences the strength of the environmental effect.
\label{fig:R3dm}}
\end{figure*}

\begin{figure*}[]
\centering\includegraphics[width=\textwidth]{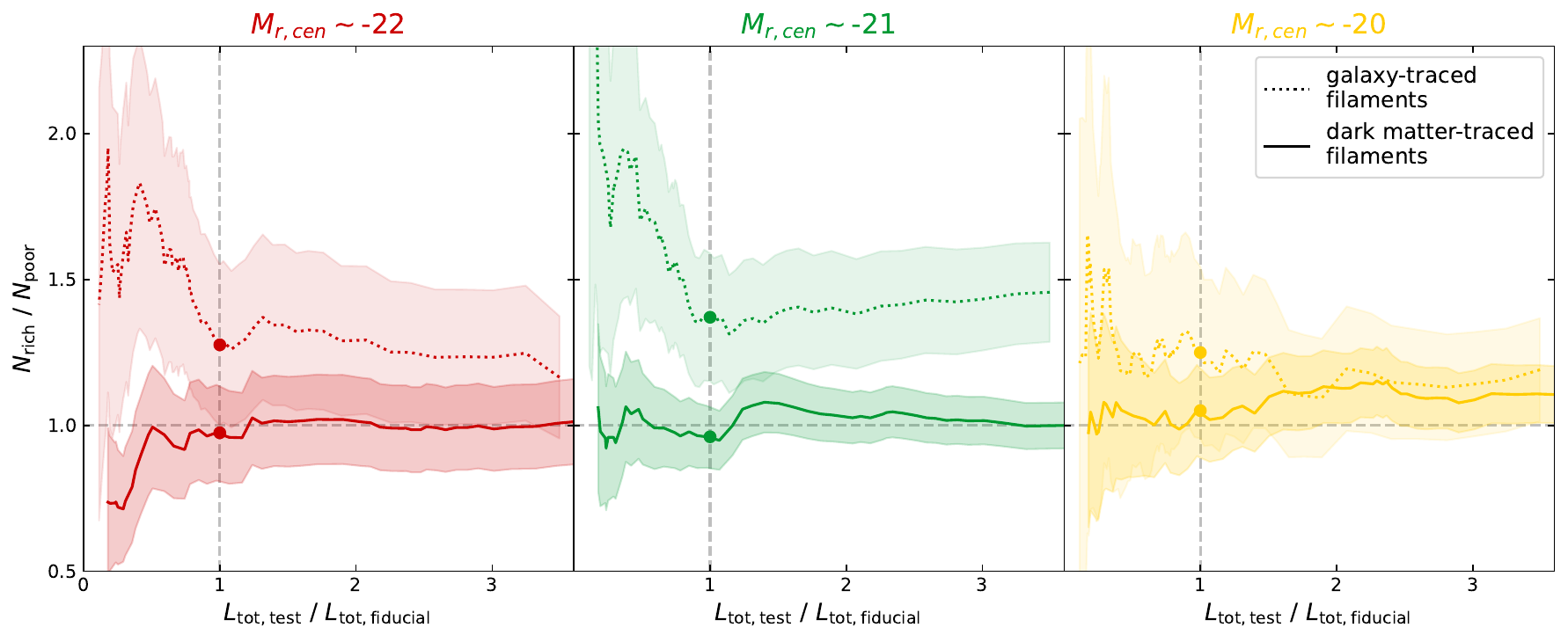}
\caption{Number ratios between satellite-rich and satellite-poor central galaxies ($N_{\rm rich}/N_{\rm poor}$) as a function of total filament length normalized to the fiducial case ($L_{\rm tot, test}/L_{\rm tot, fiducial}$; a proxy for the persistence ratio threshold). Results from different central galaxy magnitude bins are shown from left to right. Solid lines represent dark matter-traced filaments, while dotted lines correspond to galaxy-traced filaments. Shaded regions indicate uncertainties estimated from Poisson errors. The vertical gray dashed line marks fiducial parameter we choose, with markers highlighting the corresponding ratios. Across all magnitude bins and filament lengths (spanning more than an order of magnitude by varying persistence thresholds), galaxy-traced filaments consistently show a preference for hosting satellite-rich central galaxies. In contrast, dark matter-traced filaments exhibit ratios close to unity (gray dashed horizontal line), indicating no significant preference.
\label{fig:ratio}}
\end{figure*}

\subsection{Dark matter-traced filaments} \label{subsec:dm_traced_filament}

It has been shown that the choice of tracers can affect the detailed properties of filaments identified by \disperse~\citep[see e.g.,][]{2018MNRAS.474.5437L,2023MNRAS.525.4079Z,2025A&A...702A.145B}. Using cosmological simulations, these studies demonstrate that filament properties -- such as spatial distributions, lengths, and connectivity -- depend on whether dark matter particles or galaxies are used as tracers. This discrepancy arises because galaxies, distributed sparsely and discontinuously, are biased tracers of the underlying matter distribution, whereas dark matter more faithfully traces the total mass density field. In observations, we are limited to using galaxies as tracers, as was done in Section~\ref{sec:results}. In this section, we take advantage of simulations to explore the use of dark matter as tracers and investigate how the tracer choice affects the observed environmental dependence of satellite galaxy abundance.

For dark matter tracers, following \citet{2025A&A...702A.145B}, we use a smoothed density field instead of randomly sampled dark matter particles. This choice has minimal impact on the resulting filaments but yields more consistent identification with fewer artifacts at comparable computational expense. Specifically, we generate a gridded 3D density field with a cell size of 83.3 kpc, apply Gaussian smoothing with a scale of 500 kpc, and then downsample the field to a coarser grid with 250 kpc cells to reduce computational cost. Note that, since the numbers of tracer particles differ between different choices, it is usually not possible to use identical parameter sets in the \disperse~finder. Following \citet{2023MNRAS.525.4079Z}, and to enable a more direct and fair comparison with galaxy-traced filaments, we adopt a higher persistence ratio threshold of $5.1 \sigma$ (compared to $2 \sigma$ for galaxy tracers), ensuring that the total filament lengths are matched between the two cases (i.e., $L_{\rm tot, fiducial} = 7.84\times10^3$ Mpc). 
Alternatively, we have tested selecting the persistence ratio to match the number of galaxies in filaments to that of the galaxy-traced case, and found that this does not affect our main conclusions.

The resulting dark matter-traced filament spines are shown in Figure~\ref{fig:M1dm} as purple solid lines. For comparison, we also overplot the galaxy-traced filament spines using black dashed lines. The two filament networks appear qualitatively similar; however, the dark matter-traced filaments tend to be smoother and more closely follow the underlying dark matter distributions. Especially, the dark matter-traced filaments tend to extend more continuously, covering more low-density ridges that galaxy-traced filaments may miss.

\subsection{Tracer effects on satellite luminosity functions}\label{subsec:tracer_effect_LF}

Once the dark matter-traced filament spines are identified, we follow the same environmental definitions outlined in Section~\ref{subsec:env} and apply the analysis described in Section~\ref{sec:results}, as done for the galaxy-traced case. The numbers of central galaxies classified into filament and field environments across different magnitude bins are given in the two rightmost columns of Table~\ref{tab:1}. We note that the total number of central galaxies in the filament plus field environments is identical to that in the galaxy-traced case owing to our environment definition (see Section \ref{subsec:env}). Compared to the galaxy-traced case, a larger number of central galaxies are assigned to the filament environment, reflecting the fact that dark matter-traced filaments can capture more low-density ridges and thus include the galaxies surrounding these structures. The resulting environmental dependence of satellite LFs is summarized in Figure~\ref{fig:R3dm} and Table~\ref{tab:2}.

As shown in the left panels of Figure~\ref{fig:R3dm}, the ratios between the satellite LFs in filaments and in the field become smaller -- typically in the range of $1$ to $2$ -- compared to those in the galaxy-traced case, which range from $1$ to $\sim 6$. In addition, the ratios are quite similar across different magnitude bins, in contrast to the galaxy-traced case, where environmental differences become more pronounced for brighter central galaxies. When the central galaxy samples are resampled to match their $\rmagcen$ distributions (middle panels), the environmental differences decrease, similar to what is seen in the galaxy-traced case. Further resampling to match $\mvir$ distributions (right panels) leads to an even larger reduction, with the environmental differences nearly disappearing within the error bars. The average ratios are summarized in Table~\ref{tab:2}. Compared to the original environmental differences, matching both the $\rmagcen$ and $\mvir$ distributions results in relative decreases of $95 \%$, $96 \%$, and $60 \%$ in the $\rmagcen \sim -22$, $-21$, and $-20$ bins, respectively -- comparable to the results from the galaxy-traced case.

Therefore, the difference in satellite abundance in dark matter–traced filaments is less than one third of that found in galaxy-traced filaments, indicating that tracing filaments with dark matter reduces the environmental dependence of satellite LFs by more than $70 \%$. However, the relative contributions from differences in magnitude and halo mass distributions remain similar, together accounting for approximately $\sim 60 \%$ to $95 \%$ of the effect, depending on the central galaxy magnitude bin.

\subsection{How do galaxy tracers enhance the satellite abundance in filaments?}

The effects of tracer choice on satellite LFs, outlined in Section~\ref{subsec:tracer_effect_LF}, are closely related to the properties of filament-finding algorithms. Algorithms such as \disperse~ construct the cosmic web based on the spatial distribution of tracer particles. They estimate a density field from the tracer distribution and identify filaments as ridge-like structures connecting local density maxima. As a result, regions with more densely clustered tracers yield higher estimated densities and are more likely to be identified as part of the filamentary network. Consider two halos with similar virial masses residing in comparable large-scale environments -- one hosting more satellites (satellite-rich) and the other fewer (satellite-poor), although their actual surrounding dark matter densities may be similar, the satellite-rich halo contributes more to the galaxy-traced density field and is therefore more likely to be classified as residing within a galaxy-traced filament.

To illustrate this intuitively, we construct controlled central galaxy samples -- satellite-rich and satellite-poor -- from the total central galaxy sample after excluding those in the knot environment (see Section~\ref{sec:methods}). We adopt the same three central galaxy magnitude bins used previously: $\rmagcen \sim -22$, $-21$, and $-20$. In each magnitude bin, satellite-rich central galaxies are defined as those hosting at least one satellite with $M_{\rm diff} < 3$, while satellite-poor central galaxies host none.\footnote{We also tested higher $M_{\rm diff}$ thresholds to include fainter satellites (\eg~$M_{\rm diff} < 6$) and considered defining satellite-rich centrals as hosting at least one or two satellites. In all cases, the main conclusions remain consistent.}
We then further resample the satellite-rich and satellite-poor central galaxy samples to match their $\rmagcen$ and virial mass $\mvir$ distributions, ensuring a fair comparison. Finally, we examine the numbers of satellite-rich and satellite-poor central galaxies in this controlled sample that reside in the identified filaments, denoted $N_{\rm rich}$ and $N_{\rm poor}$, respectively.

For the galaxy-traced filaments, the number ratios between satellite-rich and satellite-poor central galaxies are $N_{\rm rich}/N_{\rm poor} =$ 1.28, 1.37, and 1.25 in the $\rmagcen \sim -22$, $-21$, and $-20$ bins, respectively. In contrast, these ratios are $N_{\rm rich}/N_{\rm poor} =$ 0.97, 0.96, and 1.05 in the dark matter-traced filaments (see markers in Figure~\ref{fig:ratio}).
\footnote{To ensure that difference in the magnitude distributions of central galaxies (the combined satellite-rich and satellite-poor populations) between galaxy-traced and dark matter-traced filaments does not bias the results, we performed a supplemental resampling test. After matching both samples in $\rmagcen$ and $\mvir$, the resulting $N_{\rm rich}/N_{\rm poor}$ ratios remained consistent with those reported in the main text.}
This clearly shows that, compared to the dark matter-traced case, the galaxy-traced filaments tend to pass through central galaxies hosting more satellites. This explains the more pronounced environmental differences in satellite LFs observed for the galaxy-traced case in previous subsections. Additionally, the fact that the ratio $N_{\rm rich}/N_{\rm poor}$ is quite close to $1$ in the dark matter-traced case further suggests that dark matter density is a less biased tracer of the underlying filamentary network.

Note that this result is fairly robust with respect to the \disperse~parameter choices in filament identification. For example, we have tested various persistence ratio thresholds -- which affect the total filament length -- in both tracer choices, and found the results to hold across a wide range of parameter values. In Figure~\ref{fig:ratio}, we plot the $N_{\rm rich}/N_{\rm poor}$ ratio as a function of the total filament length normalized to the fiducial case, $L_{\rm tot, test} / L_{\rm tot, fiducial}$ (a proxy for the persistence ratio threshold). Increasing the persistence threshold reduces both the number and total length of filaments, as well as the number of central galaxies residing within them. Conversely, lowering the  threshold includes more uncertain filaments, increasing the total filament length and altering the $L_{\rm tot, test} / L_{\rm tot, fiducial}$ ratio. Despite the total filament length varying by over an order of magnitude (from $L_{\rm tot, test} / L_{\rm tot, fiducial} \sim 0.1$ to $\sim 3.5$), galaxy-traced filaments consistently show a higher likelihood of including satellite-rich central galaxies. In contrast, dark matter-traced filaments yield comparable counts of satellite-rich and satellite-poor galaxies across all thresholds. This test reinforces that galaxy tracers systematically bias filament classification toward central galaxies with more satellites, amplifying the environmental dependence of satellite abundance.

\section{Summary}\label{sec:con}

We utilize the Illustris TNG100-1 hydrodynamical simulations to study the impact of cosmic filaments on satellite galaxy abundance. Filamentary structures are identified using the \disperse~ algorithm. Our main findings are summarized as follows.

(i) {\it Environmental dependence}. When filaments are identified using galaxies as tracers, we find that central galaxies in filaments tend to host more satellite galaxies than their counterparts in the field, in qualitative agreement with SDSS observational results (Figure~\ref{fig:R1}). Specifically, the average ratios between satellite LFs in filaments and those in the field are $3.49$, $2.61$, and $1.90$ in the central galaxy magnitude bins of $\rmagcen \sim -22$, $-21$, and $-20$, respectively (Table~\ref{tab:2}).

(ii) {\it Impact of magnitude and halo mass distributions}. Within each central galaxy magnitude bin, the differences in the $\rmagcen$ and host halo mass ($\mvir$) contribute significantly to the measured environmental effects. After resampling to match the $\rmagcen$ distributions in the two environments, the differences in their satellite LFs are reduced by up to $17 \%$. Further resampling to match the $\mvir$ distributions decreases the environmental differences by up to $79 \%$ (Figures~\ref{fig:R2} and \ref{fig:R3}).

(iii) {\it Impact of filament tracer choice}. The choice of filament tracers also plays a significant role in the environmental differences in satellite LFs. Compared to the galaxy-traced case, the environmental differences in the filamentary network identified using dark matter tracers are reduced by $\sim 70 \%$. However, the relative contributions from differences in magnitude and halo mass distributions remain similar, together accounting for approximately $\sim 60$--$95 \%$ of the effect on satellite abundance (Figure~\ref{fig:R3dm}).

(iv) {\it Cause of tracer effects}. The enhanced environmental effects in the galaxy-traced case is closely related to the properties of filament-finding algorithms. Compared to the dark matter density tracers, using galaxies as tracers systematically biases filament classification toward central galaxies with more satellites, amplifying the observed environmental dependence of satellite abundance (Figure~\ref{fig:ratio}).

In summary, by taking advantage of hydrodynamical simulations, we have explored the impact of halo mass distributions and filament tracer choice on the environmental dependence of satellite LFs. Our results indicate that much of the environmental differences between filament and field satellites are caused by differences in the halo mass distribution. These findings are useful for understanding and interpreting the observational results of satellite galaxy populations.

\begin{acknowledgments}
We thank the anonymous referee for their constructive comments, which helped improve this manuscript. We acknowledge support from the National Natural Science Foundation of China (NSFC) grant (No. 12588202), the National Key Program for Science and Technology Research and Development of China (2023YFB3002500), and the China Manned Space Project (CMS-CSST-2025-A03). SL acknowledges the support by the NSFC grant (No. 12473015). LW acknowledges support from the National SKA Program of China (No.2022SKA0110201), and the National Key Research and Development Program of China (No.2023YFB3002501). 
\end{acknowledgments}

\vspace{5mm}
\software{Astropy \citep{2013A&A...558A..33A,2018AJ....156..123A}, 
          \disperse~\citep{2011MNRAS.414..350S,2011MNRAS.414..384S},
          Matplotlib \citep{Hunter2007},
          Numpy \citep{Harris2020},
          Scipy \citep{Virtanen2020}.
          }

\appendix
\section{Effects of filament persistence thresholds} \label{app:filament_tests}

Filament identification with \disperse\ depends on the choice of the persistence ratio, expressed in units of $\sigma$. In the main text, we adopt a fiducial threshold of $2\sigma$. Higher thresholds select fewer filaments and filter out unreliable or spurious structures, which results in shorter total filament lengths. Conversely, lower thresholds include more filaments and lead to longer total lengths. As a result, both the number of galaxies classified as in filaments and the corresponding satellite LFs could be affected by the choice of persistence threshold.

Here we test the robustness of our conclusions using persistence thresholds of $1\sigma$, $4\sigma$, and $6\sigma$. The total filament lengths, $L_{\rm tot}$, at $1\sigma$, $4\sigma$, and $6\sigma$ are approximately $1.65$, $0.65$, and $0.26$ times the fiducial $L_{\rm tot,fiducial}$ at $2\sigma$, respectively. The persistence ratios for the dark matter–traced filaments are chosen to match the total filament lengths, as described in Section~\ref{subsec:dm_traced_filament}.

\begin{figure*}[]
\centering
\includegraphics[width=\textwidth]{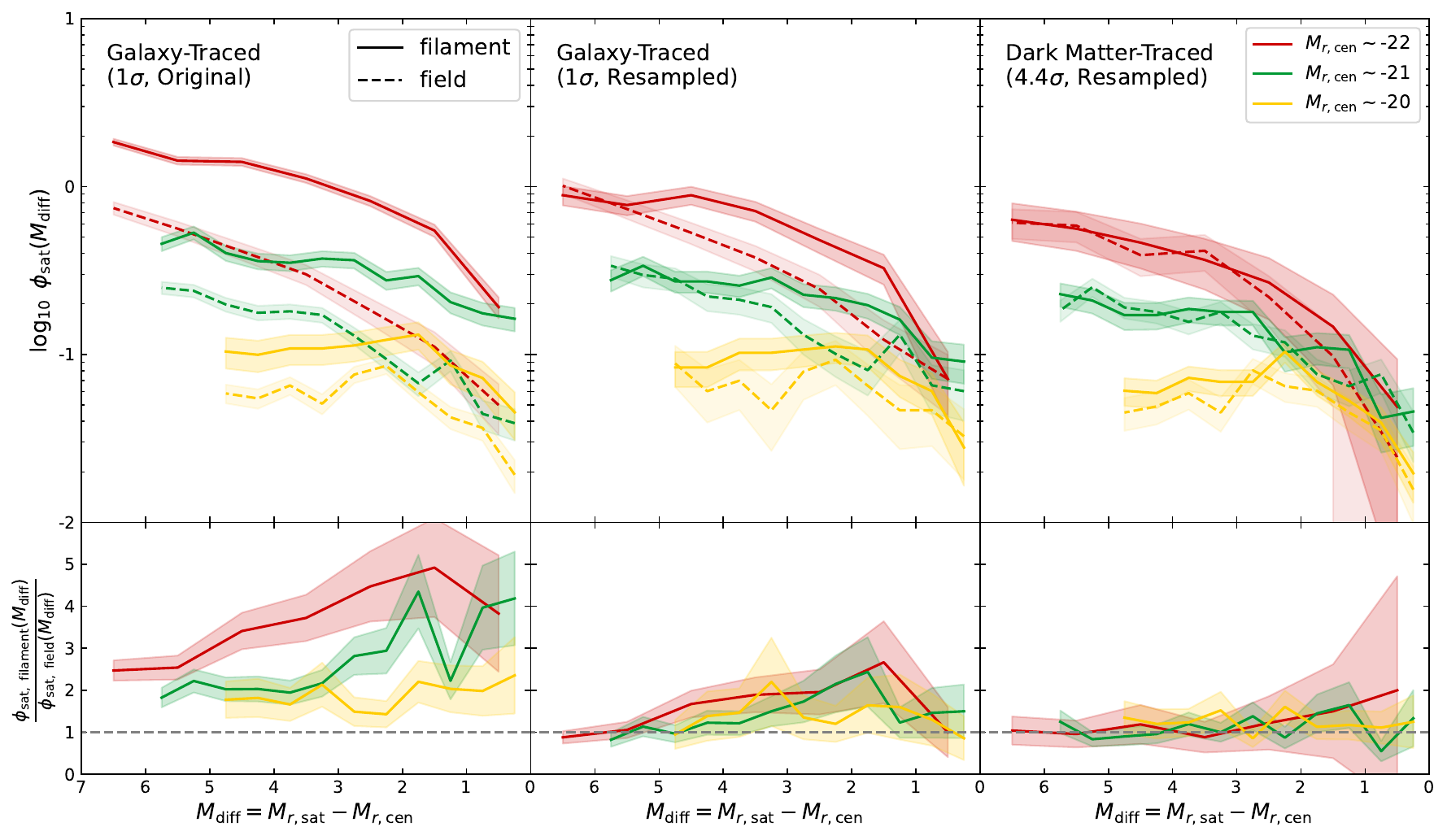}
\caption{
Environmental differences in satellite LFs, analogous to Figures~\ref{fig:R3} and \ref{fig:R3dm}, but using filaments identified with a persistence threshold of $1\sigma$. {\it Left}: Galaxy-traced filaments without resampling. {\it Middle}: Galaxy-traced filaments after resampling in $\rmagcen$ and $\mvir$. {\it Right}: Dark matter–traced filaments ($4.4\sigma$) after the same resampling. The satellite LFs and their filament-to-field ratios are quantitatively different compared to the fiducial $2\sigma$ case, but the behaviors are qualitatively similar, especially the effects of halo mass and tracer choice are consistent with the results discussed in the main text.
}
\label{fig:A1s}
\end{figure*}

\begin{figure*}[]
\centering
\includegraphics[width=\textwidth]{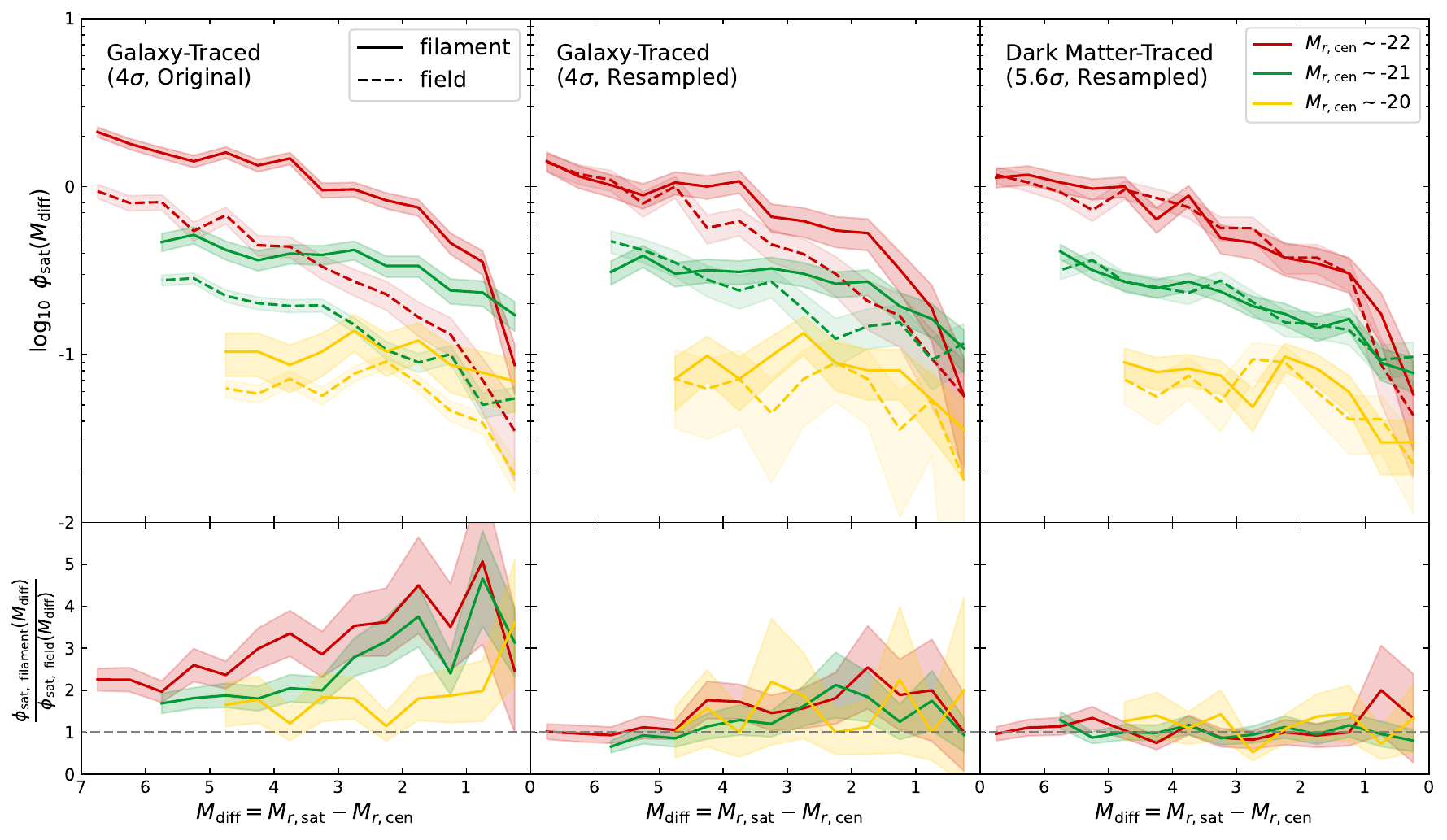}
\caption{
Same as Figure. \ref{fig:A1s}, but for filaments identifies with a higher persistence threshold of $4\sigma$.
}
\label{fig:A4s}
\end{figure*}

\begin{figure*}[]
\centering
\includegraphics[width=\textwidth]{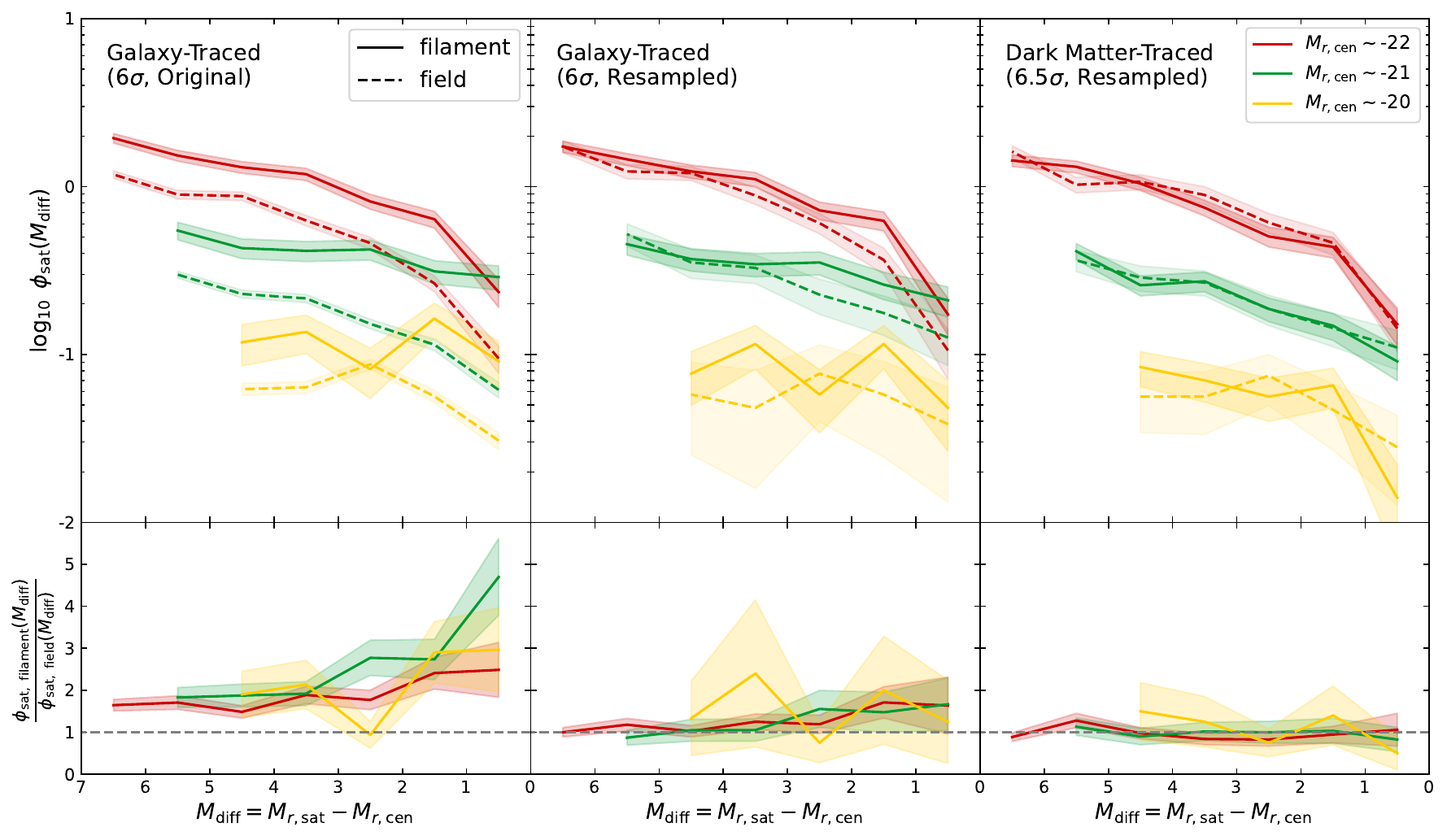}
\caption{
Same as Figure. \ref{fig:A1s}, but for filaments identifies with a higher persistence threshold of $6\sigma$.
}
\label{fig:A6s}
\end{figure*}

Figures~\ref{fig:A1s}, \ref{fig:A4s}, and \ref{fig:A6s} show the environmental differences in the satellite LFs for different persistence thresholds. Relative to Figures~\ref{fig:R3} and \ref{fig:R3dm} (the $2\sigma$ results), the satellite LFs and the ratios between filament and field environments remain broadly similar. A more noticeable deviation appears at $\rmagcen \sim -22$ (original sample, red curve) in the $6\sigma$ case, where many filaments become classified as field regions, causing the environmental ratio to drop. In all thresholds considered, resampling in $\rmagcen$ and $\mvir$ suppresses the environmental differences, and using dark matter as the filament tracer further reduces them.

For a quantitative comparison, we summarize the average ratios in Table~\ref{tab:A1}, analogous to Table~\ref{tab:2}. We find that the average satellite abundance ratio correlates with the persistence threshold: lower persistence (more filaments and longer total length) corresponds to a higher ratio, and vice versa. However, the fractional reduction from resampling remains similar across thresholds. Resampling in $\rmagcen$ reduces the ratio by $\sim 0-30\%$ (16--84th percentile), with a median reduction of $\sim 15\%$. Resampling in both $\rmagcen$ and $\mvir$ reduces the ratio by $\sim 50-90\%$ (16--84th percentile), with a median of $\sim 75\%$. For fixed total filament length, dark matter–traced filaments consistently show lower satellite abundance ratios than galaxy-traced filaments.

In summary, varying the persistence ratio from $1\sigma$ to $6\sigma$ changes the total filament length and correspondingly affects the satellite LFs and abundance ratios. Nevertheless, the main conclusions regarding environmental effects and tracer dependence remain robust.

\begin{deluxetable}{ccccc}[]
\tablecaption{
Average satellite abundance ratios between filament and field central galaxies, analogous to Table~\ref{tab:2}, but for filaments identified with different persistence thresholds ($1\sigma$, $4\sigma$, and $6\sigma$). Note that the original ratio in the last row does not exceed unity, so the calculation of the fractional reduction is not applicable and is therefore omitted.
\label{tab:A1}}
\tablehead{
    \colhead{\multirow{2}{*}{Filament tracers}} &
    \colhead{\multirow{2}{*}{Bins}} &
    \multicolumn{3}{c}{$\phi_{\rm sat, filament} / \phi_{\rm sat, field}$} \\
    \colhead{} & \colhead{} &
    \colhead{Original} &
    \colhead{Resampling $\rmagcen$} &
    \colhead{Resampling $\rmagcen$ and $\mvir$}
}
\startdata
\multirow{3}{*}{\shortstack{Galaxies \\ ($1\sigma$)}}
& $\rmagcen \sim -22$ & $3.62$ & $3.19~(17\%\downarrow)$ & $1.60~(77\%\downarrow)$ \\
& $\rmagcen \sim -21$ & $2.73$ & $2.59~(8\%\downarrow)$  & $1.44~(74\%\downarrow)$ \\
& $\rmagcen \sim -20$ & $1.89$ & $1.89~(0\%\downarrow)$  & $1.40~(56\%\downarrow)$ \\
\tableline
\multirow{3}{*}{\shortstack{Dark matter density \\ ($4.4\sigma$)}}
& $\rmagcen \sim -22$ & $3.26$ & $2.42~(37\%\downarrow)$ & $1.26~(89\%\downarrow)$ \\
& $\rmagcen \sim -21$ & $1.82$ & $1.68~(17\%\downarrow)$ & $1.11~(86\%\downarrow)$ \\
& $\rmagcen \sim -20$ & $1.64$ & $1.54~(15\%\downarrow)$ & $1.23~(63\%\downarrow)$ \\
\tableline
\multirow{3}{*}{\shortstack{Galaxies \\ ($4\sigma$)}}
& $\rmagcen \sim -22$ & $3.10$ & $2.76~(16\%\downarrow)$ & $1.50~(76\%\downarrow)$ \\
& $\rmagcen \sim -21$ & $2.59$ & $2.46~(9\%\downarrow)$  & $1.29~(82\%\downarrow)$ \\
& $\rmagcen \sim -20$ & $1.87$ & $1.99~(14\%\uparrow)$   & $1.44~(50\%\downarrow)$ \\
\tableline
\multirow{3}{*}{\shortstack{Dark matter density \\ ($5.6\sigma$)}}
& $\rmagcen \sim -22$ & $1.61$ & $1.35~(42\%\downarrow)$ & $1.09~(85\%\downarrow)$ \\
& $\rmagcen \sim -21$ & $1.36$ & $1.29~(19\%\downarrow)$ & $1.01~(98\%\downarrow)$ \\
& $\rmagcen \sim -20$ & $1.22$ & $1.21~(5\%\downarrow)$  & $1.17~(24\%\downarrow)$ \\
\tableline
\multirow{3}{*}{\shortstack{Galaxies \\ ($6\sigma$)}}
& $\rmagcen \sim -22$ & $1.86$ & $1.47~(46\%\downarrow)$ & $1.22~(75\%\downarrow)$ \\
& $\rmagcen \sim -21$ & $2.65$ & $2.43~(14\%\downarrow)$ & $1.32~(81\%\downarrow)$ \\
& $\rmagcen \sim -20$ & $2.20$ & $2.47~(22\%\uparrow)$   & $1.64~(46\%\downarrow)$ \\
\tableline
\multirow{3}{*}{\shortstack{Dark matter density \\ ($6.5\sigma$)}}
& $\rmagcen \sim -22$ & $1.37$ & $1.28~(26\%\downarrow)$ & $1.00~(100\%\downarrow)$ \\
& $\rmagcen \sim -21$ & $1.33$ & $1.30~(9\%\downarrow)$  & $1.02~(95\%\downarrow)$  \\
& $\rmagcen \sim -20$ & $0.89$ & $0.89~(-)$              & $1.08~(-)$               \\
\enddata
\end{deluxetable}

\newpage

\bibliography{references}{}

\begin{thebibliography}{}
\expandafter\ifx\csname natexlab\endcsname\relax\def\natexlab#1{#1}\fi
\providecommand{\url}[1]{\href{#1}{#1}}
\providecommand{\dodoi}[1]{doi:~\href{http://doi.org/#1}{\nolinkurl{#1}}}
\providecommand{\doeprint}[1]{\href{http://ascl.net/#1}{\nolinkurl{http://ascl.net/#1}}}
\providecommand{\doarXiv}[1]{\href{https://arxiv.org/abs/#1}{\nolinkurl{https://arxiv.org/abs/#1}}}

\bibitem[{R.~E. {Angulo} {et~al.}(2009){Angulo}, {Lacey}, {Baugh}, \& {Frenk}}]{Angulo2009}
{Angulo}, R.~E., {Lacey}, C.~G., {Baugh}, C.~M., \& {Frenk}, C.~S. 2009, \bibinfo{title}{{The fate of substructures in cold dark matter haloes},} \mnras, 399, 983, \dodoi{10.1111/j.1365-2966.2009.15333.x}

\bibitem[{M.~A. {Arag{\'o}n-Calvo} {et~al.}(2007){Arag{\'o}n-Calvo}, {van de Weygaert}, {Jones}, \& {van der Hulst}}]{2007ApJ...655L...5A}
{Arag{\'o}n-Calvo}, M.~A., {van de Weygaert}, R., {Jones}, B. J.~T., \& {van der Hulst}, J.~M. 2007, \bibinfo{title}{{Spin Alignment of Dark Matter Halos in Filaments and Walls},} \apjl, 655, L5, \dodoi{10.1086/511633}

\bibitem[{ {Astropy Collaboration} {et~al.}(2013){Astropy Collaboration}, {Robitaille}, {Tollerud}, {Greenfield}, {Droettboom}, {Bray}, {Aldcroft}, {Davis}, {Ginsburg}, {Price-Whelan}, {Kerzendorf}, {Conley}, {Crighton}, {Barbary}, {Muna}, {Ferguson}, {Grollier}, {Parikh}, {Nair}, {Unther}, {Deil}, {Woillez}, {Conseil}, {Kramer}, {Turner}, {Singer}, {Fox}, {Weaver}, {Zabalza}, {Edwards}, {Azalee Bostroem}, {Burke}, {Casey}, {Crawford}, {Dencheva}, {Ely}, {Jenness}, {Labrie}, {Lim}, {Pierfederici}, {Pontzen}, {Ptak}, {Refsdal}, {Servillat}, \& {Streicher}}]{2013A&A...558A..33A}
{Astropy Collaboration}, {Robitaille}, T.~P., {Tollerud}, E.~J., {et~al.} 2013, \bibinfo{title}{{Astropy: A community Python package for astronomy},} \aap, 558, A33, \dodoi{10.1051/0004-6361/201322068}

\bibitem[{ {Astropy Collaboration} {et~al.}(2018){Astropy Collaboration}, {Price-Whelan}, {Sip{\H{o}}cz}, {G{\"u}nther}, {Lim}, {Crawford}, {Conseil}, {Shupe}, {Craig}, {Dencheva}, {Ginsburg}, {VanderPlas}, {Bradley}, {P{\'e}rez-Su{\'a}rez}, {de Val-Borro}, {Aldcroft}, {Cruz}, {Robitaille}, {Tollerud}, {Ardelean}, {Babej}, {Bach}, {Bachetti}, {Bakanov}, {Bamford}, {Barentsen}, {Barmby}, {Baumbach}, {Berry}, {Biscani}, {Boquien}, {Bostroem}, {Bouma}, {Brammer}, {Bray}, {Breytenbach}, {Buddelmeijer}, {Burke}, {Calderone}, {Cano Rodr{\'\i}guez}, {Cara}, {Cardoso}, {Cheedella}, {Copin}, {Corrales}, {Crichton}, {D'Avella}, {Deil}, {Depagne}, {Dietrich}, {Donath}, {Droettboom}, {Earl}, {Erben}, {Fabbro}, {Ferreira}, {Finethy}, {Fox}, {Garrison}, {Gibbons}, {Goldstein}, {Gommers}, {Greco}, {Greenfield}, {Groener}, {Grollier}, {Hagen}, {Hirst}, {Homeier}, {Horton}, {Hosseinzadeh}, {Hu}, {Hunkeler}, {Ivezi{\'c}}, {Jain}, {Jenness}, {Kanarek}, {Kendrew}, {Kern}, {Kerzendorf}, {Khvalko}, {King}, {Kirkby}, {Kulkarni},
  {Kumar}, {Lee}, {Lenz}, {Littlefair}, {Ma}, {Macleod}, {Mastropietro}, {McCully}, {Montagnac}, {Morris}, {Mueller}, {Mumford}, {Muna}, {Murphy}, {Nelson}, {Nguyen}, {Ninan}, {N{\"o}the}, {Ogaz}, {Oh}, {Parejko}, {Parley}, {Pascual}, {Patil}, {Patil}, {Plunkett}, {Prochaska}, {Rastogi}, {Reddy Janga}, {Sabater}, {Sakurikar}, {Seifert}, {Sherbert}, {Sherwood-Taylor}, {Shih}, {Sick}, {Silbiger}, {Singanamalla}, {Singer}, {Sladen}, {Sooley}, {Sornarajah}, {Streicher}, {Teuben}, {Thomas}, {Tremblay}, {Turner}, {Terr{\'o}n}, {van Kerkwijk}, {de la Vega}, {Watkins}, {Weaver}, {Whitmore}, {Woillez}, {Zabalza}, \& {Astropy Contributors}}]{2018AJ....156..123A}
{Astropy Collaboration}, {Price-Whelan}, A.~M., {Sip{\H{o}}cz}, B.~M., {et~al.} 2018, \bibinfo{title}{{The Astropy Project: Building an Open-science Project and Status of the v2.0 Core Package},} \aj, 156, 123, \dodoi{10.3847/1538-3881/aabc4f}

\bibitem[{Y.~M. {Bah{\'e}} \& P. {Jablonka}(2025){Bah{\'e}} \& {Jablonka}}]{2025A&A...702A.145B}
{Bah{\'e}}, Y.~M., \& {Jablonka}, P. 2025, \bibinfo{title}{{Galaxies in the simulated cosmic web: I. Filament identification and properties},} \aap, 702, A145, \dodoi{10.1051/0004-6361/202554079}

\bibitem[{M.~R. {Blanton} {et~al.}(2005){Blanton}, {Eisenstein}, {Hogg}, {Schlegel}, \& {Brinkmann}}]{2005ApJ...629..143B}
{Blanton}, M.~R., {Eisenstein}, D., {Hogg}, D.~W., {Schlegel}, D.~J., \& {Brinkmann}, J. 2005, \bibinfo{title}{{Relationship between Environment and the Broadband Optical Properties of Galaxies in the Sloan Digital Sky Survey},} \apj, 629, 143, \dodoi{10.1086/422897}

\bibitem[{J.~R. {Bond} {et~al.}(1996){Bond}, {Kofman}, \& {Pogosyan}}]{Bond1996}
{Bond}, J.~R., {Kofman}, L., \& {Pogosyan}, D. 1996, \bibinfo{title}{{How filaments of galaxies are woven into the cosmic web},} \nat, 380, 603, \dodoi{10.1038/380603a0}

\bibitem[{M. {Boylan-Kolchin} {et~al.}(2010){Boylan-Kolchin}, {Springel}, {White}, \& {Jenkins}}]{Boylan-Kolchin2010}
{Boylan-Kolchin}, M., {Springel}, V., {White}, S. D.~M., \& {Jenkins}, A. 2010, \bibinfo{title}{{There's no place like home? Statistics of Milky Way-mass dark matter haloes},} \mnras, 406, 896, \dodoi{10.1111/j.1365-2966.2010.16774.x}

\bibitem[{J.~S. {Bullock} \& M. {Boylan-Kolchin}(2017){Bullock} \& {Boylan-Kolchin}}]{Bullock2017}
{Bullock}, J.~S., \& {Boylan-Kolchin}, M. 2017, \bibinfo{title}{{Small-Scale Challenges to the {\ensuremath{\Lambda}}CDM Paradigm},} \araa, 55, 343, \dodoi{10.1146/annurev-astro-091916-055313}

\bibitem[{G. {Castignani} {et~al.}(2022){Castignani}, {Combes}, {Jablonka}, {Finn}, {Rudnick}, {Vulcani}, {Desai}, {Zaritsky}, \& {Salom{\'e}}}]{2022A&A...657A...9C}
{Castignani}, G., {Combes}, F., {Jablonka}, P., {et~al.} 2022, \bibinfo{title}{{Virgo filaments. I. Processing of gas in cosmological filaments around the Virgo cluster},} \aap, 657, A9, \dodoi{10.1051/0004-6361/202040141}

\bibitem[{M. {Cautun} {et~al.}(2014){Cautun}, {van de Weygaert}, {Jones}, \& {Frenk}}]{Cautun2014}
{Cautun}, M., {van de Weygaert}, R., {Jones}, B. J.~T., \& {Frenk}, C.~S. 2014, \bibinfo{title}{{Evolution of the cosmic web},} \mnras, 441, 2923, \dodoi{10.1093/mnras/stu768}

\bibitem[{S. {Chandrasekhar}(1943){Chandrasekhar}}]{1943ApJ....97..255C}
{Chandrasekhar}, S. 1943, \bibinfo{title}{{Dynamical Friction. I. General Considerations: the Coefficient of Dynamical Friction.},} \apj, 97, 255, \dodoi{10.1086/144517}

\bibitem[{S. {Codis} {et~al.}(2018){Codis}, {Jindal}, {Chisari}, {Vibert}, {Dubois}, {Pichon}, \& {Devriendt}}]{Codis2018}
{Codis}, S., {Jindal}, A., {Chisari}, N.~E., {et~al.} 2018, \bibinfo{title}{{Galaxy orientation with the cosmic web across cosmic time},} \mnras, 481, 4753, \dodoi{10.1093/mnras/sty2567}

\bibitem[{R.~A. {Crain} {et~al.}(2009){Crain}, {Theuns}, {Dalla Vecchia}, {Eke}, {Frenk}, {Jenkins}, {Kay}, {Peacock}, {Pearce}, {Schaye}, {Springel}, {Thomas}, {White}, \& {Wiersma}}]{Crain2009}
{Crain}, R.~A., {Theuns}, T., {Dalla Vecchia}, C., {et~al.} 2009, \bibinfo{title}{{Galaxies-intergalactic medium interaction calculation - I. Galaxy formation as a function of large-scale environment},} \mnras, 399, 1773, \dodoi{10.1111/j.1365-2966.2009.15402.x}

\bibitem[{B. {Darvish} {et~al.}(2017){Darvish}, {Mobasher}, {Martin}, {Sobral}, {Scoville}, {Stroe}, {Hemmati}, \& {Kartaltepe}}]{2017ApJ...837...16D}
{Darvish}, B., {Mobasher}, B., {Martin}, D.~C., {et~al.} 2017, \bibinfo{title}{{Cosmic Web of Galaxies in the COSMOS Field: Public Catalog and Different Quenching for Centrals and Satellites},} \apj, 837, 16, \dodoi{10.3847/1538-4357/837/1/16}

\bibitem[{M. {Davis} {et~al.}(1985){Davis}, {Efstathiou}, {Frenk}, \& {White}}]{Davis1985}
{Davis}, M., {Efstathiou}, G., {Frenk}, C.~S., \& {White}, S.~D.~M. 1985, \bibinfo{title}{{The evolution of large-scale structure in a universe dominated by cold dark matter},} \apj, 292, 371, \dodoi{10.1086/163168}

\bibitem[{J. {Diemand} {et~al.}(2007){Diemand}, {Kuhlen}, \& {Madau}}]{Diemand2007}
{Diemand}, J., {Kuhlen}, M., \& {Madau}, P. 2007, \bibinfo{title}{{Formation and Evolution of Galaxy Dark Matter Halos and Their Substructure},} \apj, 667, 859, \dodoi{10.1086/520573}

\bibitem[{K. {Dolag} {et~al.}(2009){Dolag}, {Borgani}, {Murante}, \& {Springel}}]{Dolag2009}
{Dolag}, K., {Borgani}, S., {Murante}, G., \& {Springel}, V. 2009, \bibinfo{title}{{Substructures in hydrodynamical cluster simulations},} \mnras, 399, 497, \dodoi{10.1111/j.1365-2966.2009.15034.x}

\bibitem[{Y. {Dubois} {et~al.}(2014){Dubois}, {Pichon}, {Welker}, {Le Borgne}, {Devriendt}, {Laigle}, {Codis}, {Pogosyan}, {Arnouts}, {Benabed}, {Bertin}, {Blaizot}, {Bouchet}, {Cardoso}, {Colombi}, {de Lapparent}, {Desjacques}, {Gavazzi}, {Kassin}, {Kimm}, {McCracken}, {Milliard}, {Peirani}, {Prunet}, {Rouberol}, {Silk}, {Slyz}, {Sousbie}, {Teyssier}, {Tresse}, {Treyer}, {Vibert}, \& {Volonteri}}]{Dubois2014}
{Dubois}, Y., {Pichon}, C., {Welker}, C., {et~al.} 2014, \bibinfo{title}{{Dancing in the dark: galactic properties trace spin swings along the cosmic web},} \mnras, 444, 1453, \dodoi{10.1093/mnras/stu1227}

\bibitem[{O. {Fakhouri} \& C.-P. {Ma}(2009){Fakhouri} \& {Ma}}]{2009MNRAS.394.1825F}
{Fakhouri}, O., \& {Ma}, C.-P. 2009, \bibinfo{title}{{Environmental dependence of dark matter halo growth - I. Halo merger rates},} \mnras, 394, 1825, \dodoi{10.1111/j.1365-2966.2009.14480.x}

\bibitem[{C.~S. {Frenk} \& S.~D.~M. {White}(2012){Frenk} \& {White}}]{Frenk2012}
{Frenk}, C.~S., \& {White}, S.~D.~M. 2012, \bibinfo{title}{{Dark matter and cosmic structure},} Annalen der Physik, 524, 507, \dodoi{10.1002/andp.201200212}

\bibitem[{D. {Gal{\'a}rraga-Espinosa} {et~al.}(2024){Gal{\'a}rraga-Espinosa}, {Cadiou}, {Gouin}, {White}, {Springel}, {Pakmor}, {Hadzhiyska}, {Bose}, {Ferlito}, {Hernquist}, {Kannan}, {Barrera}, {Maria Delgado}, \& {Hern{\'a}ndez-Aguayo}}]{2024A&A...684A..63G}
{Gal{\'a}rraga-Espinosa}, D., {Cadiou}, C., {Gouin}, C., {et~al.} 2024, \bibinfo{title}{{Evolution of cosmic filaments in the MTNG simulation},} \aap, 684, A63, \dodoi{10.1051/0004-6361/202347982}

\bibitem[{P. {Ganeshaiah Veena} {et~al.}(2019){Ganeshaiah Veena}, {Cautun}, {Tempel}, {van de Weygaert}, \& {Frenk}}]{GaneshaiahVeena2019}
{Ganeshaiah Veena}, P., {Cautun}, M., {Tempel}, E., {van de Weygaert}, R., \& {Frenk}, C.~S. 2019, \bibinfo{title}{{The Cosmic Ballet II: spin alignment of galaxies and haloes with large-scale filaments in the EAGLE simulation},} \mnras, 487, 1607, \dodoi{10.1093/mnras/stz1343}

\bibitem[{L. {Gao} {et~al.}(2011){Gao}, {Frenk}, {Boylan-Kolchin}, {Jenkins}, {Springel}, \& {White}}]{Gao2011}
{Gao}, L., {Frenk}, C.~S., {Boylan-Kolchin}, M., {et~al.} 2011, \bibinfo{title}{{The statistics of the subhalo abundance of dark matter haloes},} \mnras, 410, 2309, \dodoi{10.1111/j.1365-2966.2010.17601.x}

\bibitem[{L. {Gao} {et~al.}(2004){Gao}, {White}, {Jenkins}, {Stoehr}, \& {Springel}}]{Gao2004}
{Gao}, L., {White}, S.~D.~M., {Jenkins}, A., {Stoehr}, F., \& {Springel}, V. 2004, \bibinfo{title}{{The subhalo populations of {\ensuremath{\Lambda}}CDM dark haloes},} \mnras, 355, 819, \dodoi{10.1111/j.1365-2966.2004.08360.x}

\bibitem[{Q. {Guo} {et~al.}(2015){Guo}, {Tempel}, \& {Libeskind}}]{2015ApJ...800..112G}
{Guo}, Q., {Tempel}, E., \& {Libeskind}, N.~I. 2015, \bibinfo{title}{{Galaxies in Filaments have More Satellites: The Influence of the Cosmic Web on the Satellite Luminosity Function in the SDSS},} \apj, 800, 112, \dodoi{10.1088/0004-637X/800/2/112}

\bibitem[{O. {Hahn} {et~al.}(2007{\natexlab{a}}){Hahn}, {Carollo}, {Porciani}, \& {Dekel}}]{2007MNRAS.381...41H}
{Hahn}, O., {Carollo}, C.~M., {Porciani}, C., \& {Dekel}, A. 2007{\natexlab{a}}, \bibinfo{title}{{The evolution of dark matter halo properties in clusters, filaments, sheets and voids},} \mnras, 381, 41, \dodoi{10.1111/j.1365-2966.2007.12249.x}

\bibitem[{O. {Hahn} {et~al.}(2007{\natexlab{b}}){Hahn}, {Porciani}, {Carollo}, \& {Dekel}}]{2007MNRAS.375..489H}
{Hahn}, O., {Porciani}, C., {Carollo}, C.~M., \& {Dekel}, A. 2007{\natexlab{b}}, \bibinfo{title}{{Properties of dark matter haloes in clusters, filaments, sheets and voids},} \mnras, 375, 489, \dodoi{10.1111/j.1365-2966.2006.11318.x}

\bibitem[{C.~R. {Harris} {et~al.}(2020){Harris}, {Millman}, {van der Walt}, {Gommers}, {Virtanen}, {Cournapeau}, {Wieser}, {Taylor}, {Berg}, {Smith}, {Kern}, {Picus}, {Hoyer}, {van Kerkwijk}, {Brett}, {Haldane}, {del R{\'\i}o}, {Wiebe}, {Peterson}, {G{\'e}rard-Marchant}, {Sheppard}, {Reddy}, {Weckesser}, {Abbasi}, {Gohlke}, \& {Oliphant}}]{Harris2020}
{Harris}, C.~R., {Millman}, K.~J., {van der Walt}, S.~J., {et~al.} 2020, \bibinfo{title}{{Array programming with NumPy},} \nat, 585, 357, \dodoi{10.1038/s41586-020-2649-2}

\bibitem[{M. {Hoosain} {et~al.}(2024){Hoosain}, {Blyth}, {Skelton}, {Kannappan}, {Stark}, {Eckert}, {Hutchens}, {Carr}, \& {Kraljic}}]{2024MNRAS.528.4139H}
{Hoosain}, M., {Blyth}, S.-L., {Skelton}, R.~E., {et~al.} 2024, \bibinfo{title}{{The effect of cosmic web filaments on galaxy properties in the RESOLVE and ECO surveys},} \mnras, 528, 4139, \dodoi{10.1093/mnras/stae174}

\bibitem[{J.~D. {Hunter}(2007){Hunter}}]{Hunter2007}
{Hunter}, J.~D. 2007, \bibinfo{title}{{Matplotlib: A 2D Graphics Environment},} Computing in Science and Engineering, 9, 90, \dodoi{10.1109/MCSE.2007.55}

\bibitem[{F. {Jiang} \& F.~C. {van den Bosch}(2017){Jiang} \& {van den Bosch}}]{Jiang2017}
{Jiang}, F., \& {van den Bosch}, F.~C. 2017, \bibinfo{title}{{Statistics of dark matter substructure - III. Halo-to-halo variance},} \mnras, 472, 657, \dodoi{10.1093/mnras/stx1979}

\bibitem[{G. {Kauffmann} {et~al.}(2004){Kauffmann}, {White}, {Heckman}, {M{\'e}nard}, {Brinchmann}, {Charlot}, {Tremonti}, \& {Brinkmann}}]{2004MNRAS.353..713K}
{Kauffmann}, G., {White}, S. D.~M., {Heckman}, T.~M., {et~al.} 2004, \bibinfo{title}{{The environmental dependence of the relations between stellar mass, structure, star formation and nuclear activity in galaxies},} \mnras, 353, 713, \dodoi{10.1111/j.1365-2966.2004.08117.x}

\bibitem[{A.~A. {Klypin} {et~al.}(2011){Klypin}, {Trujillo-Gomez}, \& {Primack}}]{Klypin2011}
{Klypin}, A.~A., {Trujillo-Gomez}, S., \& {Primack}, J. 2011, \bibinfo{title}{{Dark Matter Halos in the Standard Cosmological Model: Results from the Bolshoi Simulation},} \apj, 740, 102, \dodoi{10.1088/0004-637X/740/2/102}

\bibitem[{C. {Laigle} {et~al.}(2018){Laigle}, {Pichon}, {Arnouts}, {McCracken}, {Dubois}, {Devriendt}, {Slyz}, {Le Borgne}, {Benoit-L{\'e}vy}, {Hwang}, {Ilbert}, {Kraljic}, {Malavasi}, {Park}, \& {Vibert}}]{2018MNRAS.474.5437L}
{Laigle}, C., {Pichon}, C., {Arnouts}, S., {et~al.} 2018, \bibinfo{title}{{COSMOS2015 photometric redshifts probe the impact of filaments on galaxy properties},} \mnras, 474, 5437, \dodoi{10.1093/mnras/stx3055}

\bibitem[{S. {Liao} \& L. {Gao}(2019){Liao} \& {Gao}}]{2019MNRAS.485..464L}
{Liao}, S., \& {Gao}, L. 2019, \bibinfo{title}{{Impact of filaments on galaxy formation in their residing dark matter haloes},} \mnras, 485, 464, \dodoi{10.1093/mnras/stz441}

\bibitem[{N.~I. {Libeskind} {et~al.}(2012){Libeskind}, {Hoffman}, {Knebe}, {Steinmetz}, {Gottl{\"o}ber}, {Metuki}, \& {Yepes}}]{2012MNRAS.421L.137L}
{Libeskind}, N.~I., {Hoffman}, Y., {Knebe}, A., {et~al.} 2012, \bibinfo{title}{{The cosmic web and the orientation of angular momenta},} \mnras, 421, L137, \dodoi{10.1111/j.1745-3933.2012.01222.x}

\bibitem[{N.~I. {Libeskind} {et~al.}(2018){Libeskind}, {van de Weygaert}, {Cautun}, {Falck}, {Tempel}, {Abel}, {Alpaslan}, {Arag{\'o}n-Calvo}, {Forero-Romero}, {Gonzalez}, {Gottl{\"o}ber}, {Hahn}, {Hellwing}, {Hoffman}, {Jones}, {Kitaura}, {Knebe}, {Manti}, {Neyrinck}, {Nuza}, {Padilla}, {Platen}, {Ramachandra}, {Robotham}, {Saar}, {Shandarin}, {Steinmetz}, {Stoica}, {Sousbie}, \& {Yepes}}]{2018MNRAS.473.1195L}
{Libeskind}, N.~I., {van de Weygaert}, R., {Cautun}, M., {et~al.} 2018, \bibinfo{title}{{Tracing the cosmic web},} \mnras, 473, 1195, \dodoi{10.1093/mnras/stx1976}

\bibitem[{Y. {Liu} {et~al.}(2025){Liu}, {Gao}, {Liao}, \& {Zhu}}]{Liu2025}
{Liu}, Y., {Gao}, L., {Liao}, S., \& {Zhu}, K. 2025, \bibinfo{title}{{Prospects for Detecting Cosmic Filaments in Ly{\ensuremath{\alpha}} Emission across Redshifts z = 2{\textendash}5},} \apj, 984, 55, \dodoi{10.3847/1538-4357/adc44b}

\bibitem[{N. {Malavasi} {et~al.}(2020{\natexlab{a}}){Malavasi}, {Aghanim}, {Douspis}, {Tanimura}, \& {Bonjean}}]{2020A&A...642A..19M}
{Malavasi}, N., {Aghanim}, N., {Douspis}, M., {Tanimura}, H., \& {Bonjean}, V. 2020{\natexlab{a}}, \bibinfo{title}{{Characterising filaments in the SDSS volume from the galaxy distribution},} \aap, 642, A19, \dodoi{10.1051/0004-6361/202037647}

\bibitem[{N. {Malavasi} {et~al.}(2020{\natexlab{b}}){Malavasi}, {Aghanim}, {Tanimura}, {Bonjean}, \& {Douspis}}]{2020A&A...634A..30M}
{Malavasi}, N., {Aghanim}, N., {Tanimura}, H., {Bonjean}, V., \& {Douspis}, M. 2020{\natexlab{b}}, \bibinfo{title}{{Like a spider in its web: a study of the large-scale structure around the Coma cluster},} \aap, 634, A30, \dodoi{10.1051/0004-6361/201936629}

\bibitem[{F. {Marinacci} {et~al.}(2018){Marinacci}, {Vogelsberger}, {Pakmor}, {Torrey}, {Springel}, {Hernquist}, {Nelson}, {Weinberger}, {Pillepich}, {Naiman}, \& {Genel}}]{2018MNRAS.480.5113M}
{Marinacci}, F., {Vogelsberger}, M., {Pakmor}, R., {et~al.} 2018, \bibinfo{title}{{First results from the IllustrisTNG simulations: radio haloes and magnetic fields},} \mnras, 480, 5113, \dodoi{10.1093/mnras/sty2206}

\bibitem[{F. {Markos Hunde} {et~al.}(2025){Markos Hunde}, {Newton}, {Hellwing}, {Bilicki}, \& {Naidoo}}]{2025A&A...700A..65M}
{Markos Hunde}, F., {Newton}, O., {Hellwing}, W.~A., {Bilicki}, M., \& {Naidoo}, K. 2025, \bibinfo{title}{{Caught in the cosmic web: Environmental effects on subhalo abundance and internal density profiles},} \aap, 700, A65, \dodoi{10.1051/0004-6361/202452246}

\bibitem[{O. {Metuki} {et~al.}(2015){Metuki}, {Libeskind}, {Hoffman}, {Crain}, \& {Theuns}}]{2015MNRAS.446.1458M}
{Metuki}, O., {Libeskind}, N.~I., {Hoffman}, Y., {Crain}, R.~A., \& {Theuns}, T. 2015, \bibinfo{title}{{Galaxy properties and the cosmic web in simulations},} \mnras, 446, 1458, \dodoi{10.1093/mnras/stu2166}

\bibitem[{H. {Mo} {et~al.}(2010){Mo}, {van den Bosch}, \& {White}}]{Mo2010}
{Mo}, H., {van den Bosch}, F.~C., \& {White}, S. 2010, {Galaxy Formation and Evolution} ({Cambridge, UK}: {Cambridge University Press})

\bibitem[{J.~P. {Naiman} {et~al.}(2018){Naiman}, {Pillepich}, {Springel}, {Ramirez-Ruiz}, {Torrey}, {Vogelsberger}, {Pakmor}, {Nelson}, {Marinacci}, {Hernquist}, {Weinberger}, \& {Genel}}]{2018MNRAS.477.1206N}
{Naiman}, J.~P., {Pillepich}, A., {Springel}, V., {et~al.} 2018, \bibinfo{title}{{First results from the IllustrisTNG simulations: a tale of two elements - chemical evolution of magnesium and europium},} \mnras, 477, 1206, \dodoi{10.1093/mnras/sty618}

\bibitem[{A. {Nandi} \& B. {Pandey}(2025){Nandi} \& {Pandey}}]{2025arXiv250718614N}
{Nandi}, A., \& {Pandey}, B. 2025, \bibinfo{title}{{Galaxy quenching across the Cosmic Web: disentangling mass and environment with SDSS DR18},} arXiv e-prints, arXiv:2507.18614, \dodoi{10.48550/arXiv.2507.18614}

\bibitem[{D. {Nelson} {et~al.}(2018){Nelson}, {Pillepich}, {Springel}, {Weinberger}, {Hernquist}, {Pakmor}, {Genel}, {Torrey}, {Vogelsberger}, {Kauffmann}, {Marinacci}, \& {Naiman}}]{2018MNRAS.475..624N}
{Nelson}, D., {Pillepich}, A., {Springel}, V., {et~al.} 2018, \bibinfo{title}{{First results from the IllustrisTNG simulations: the galaxy colour bimodality},} \mnras, 475, 624, \dodoi{10.1093/mnras/stx3040}

\bibitem[{D. {Nelson} {et~al.}(2019){Nelson}, {Springel}, {Pillepich}, {Rodriguez-Gomez}, {Torrey}, {Genel}, {Vogelsberger}, {Pakmor}, {Marinacci}, {Weinberger}, {Kelley}, {Lovell}, {Diemer}, \& {Hernquist}}]{2019ComAC...6....2N}
{Nelson}, D., {Springel}, V., {Pillepich}, A., {et~al.} 2019, \bibinfo{title}{{The IllustrisTNG simulations: public data release},} Computational Astrophysics and Cosmology, 6, 2, \dodoi{10.1186/s40668-019-0028-x}

\bibitem[{A. {Pillepich} {et~al.}(2018{\natexlab{a}}){Pillepich}, {Nelson}, {Hernquist}, {Springel}, {Pakmor}, {Torrey}, {Weinberger}, {Genel}, {Naiman}, {Marinacci}, \& {Vogelsberger}}]{Pillepich2018}
{Pillepich}, A., {Nelson}, D., {Hernquist}, L., {et~al.} 2018{\natexlab{a}}, \bibinfo{title}{{First results from the IllustrisTNG simulations: the stellar mass content of groups and clusters of galaxies},} \mnras, 475, 648, \dodoi{10.1093/mnras/stx3112}

\bibitem[{A. {Pillepich} {et~al.}(2018{\natexlab{b}}){Pillepich}, {Springel}, {Nelson}, {Genel}, {Naiman}, {Pakmor}, {Hernquist}, {Torrey}, {Vogelsberger}, {Weinberger}, \& {Marinacci}}]{2018MNRAS.473.4077P}
{Pillepich}, A., {Springel}, V., {Nelson}, D., {et~al.} 2018{\natexlab{b}}, \bibinfo{title}{{Simulating galaxy formation with the IllustrisTNG model},} \mnras, 473, 4077, \dodoi{10.1093/mnras/stx2656}

\bibitem[{ {Planck Collaboration} {et~al.}(2016){Planck Collaboration}, {Ade}, {Aghanim}, {Arnaud}, {Ashdown}, {Aumont}, {Baccigalupi}, {Banday}, {Barreiro}, {Bartlett}, {Bartolo}, {Battaner}, {Battye}, {Benabed}, {Beno{\^\i}t}, {Benoit-L{\'e}vy}, {Bernard}, {Bersanelli}, {Bielewicz}, {Bock}, {Bonaldi}, {Bonavera}, {Bond}, {Borrill}, {Bouchet}, {Boulanger}, {Bucher}, {Burigana}, {Butler}, {Calabrese}, {Cardoso}, {Catalano}, {Challinor}, {Chamballu}, {Chary}, {Chiang}, {Chluba}, {Christensen}, {Church}, {Clements}, {Colombi}, {Colombo}, {Combet}, {Coulais}, {Crill}, {Curto}, {Cuttaia}, {Danese}, {Davies}, {Davis}, {de Bernardis}, {de Rosa}, {de Zotti}, {Delabrouille}, {D{\'e}sert}, {Di Valentino}, {Dickinson}, {Diego}, {Dolag}, {Dole}, {Donzelli}, {Dor{\'e}}, {Douspis}, {Ducout}, {Dunkley}, {Dupac}, {Efstathiou}, {Elsner}, {En{\ss}lin}, {Eriksen}, {Farhang}, {Fergusson}, {Finelli}, {Forni}, {Frailis}, {Fraisse}, {Franceschi}, {Frejsel}, {Galeotta}, {Galli}, {Ganga}, {Gauthier}, {Gerbino}, {Ghosh}, {Giard},
  {Giraud-H{\'e}raud}, {Giusarma}, {Gjerl{\o}w}, {Gonz{\'a}lez-Nuevo}, {G{\'o}rski}, {Gratton}, {Gregorio}, {Gruppuso}, {Gudmundsson}, {Hamann}, {Hansen}, {Hanson}, {Harrison}, {Helou}, {Henrot-Versill{\'e}}, {Hern{\'a}ndez-Monteagudo}, {Herranz}, {Hildebrandt}, {Hivon}, {Hobson}, {Holmes}, {Hornstrup}, {Hovest}, {Huang}, {Huffenberger}, {Hurier}, {Jaffe}, {Jaffe}, {Jones}, {Juvela}, {Keih{\"a}nen}, {Keskitalo}, {Kisner}, {Kneissl}, {Knoche}, {Knox}, {Kunz}, {Kurki-Suonio}, {Lagache}, {L{\"a}hteenm{\"a}ki}, {Lamarre}, {Lasenby}, {Lattanzi}, {Lawrence}, {Leahy}, {Leonardi}, {Lesgourgues}, {Levrier}, {Lewis}, {Liguori}, {Lilje}, {Linden-V{\o}rnle}, {L{\'o}pez-Caniego}, {Lubin}, {Mac{\'\i}as-P{\'e}rez}, {Maggio}, {Maino}, {Mandolesi}, {Mangilli}, {Marchini}, {Maris}, {Martin}, {Martinelli}, {Mart{\'\i}nez-Gonz{\'a}lez}, {Masi}, {Matarrese}, {McGehee}, {Meinhold}, {Melchiorri}, {Melin}, {Mendes}, {Mennella}, {Migliaccio}, {Millea}, {Mitra}, {Miville-Desch{\^e}nes}, {Moneti}, {Montier}, {Morgante}, {Mortlock},
  {Moss}, {Munshi}, {Murphy}, {Naselsky}, {Nati}, {Natoli}, {Netterfield}, {N{\o}rgaard-Nielsen}, {Noviello}, {Novikov}, {Novikov}, {Oxborrow}, {Paci}, {Pagano}, {Pajot}, {Paladini}, {Paoletti}, {Partridge}, {Pasian}, {Patanchon}, {Pearson}, {Perdereau}, {Perotto}, {Perrotta}, {Pettorino}, {Piacentini}, {Piat}, {Pierpaoli}, {Pietrobon}, {Plaszczynski}, {Pointecouteau}, {Polenta}, {Popa}, {Pratt}, {Pr{\'e}zeau}, {Prunet}, {Puget}, {Rachen}, {Reach}, {Rebolo}, {Reinecke}, {Remazeilles}, {Renault}, {Renzi}, {Ristorcelli}, {Rocha}, {Rosset}, {Rossetti}, {Roudier}, {Rouill{\'e} d'Orfeuil}, {Rowan-Robinson}, {Rubi{\~n}o-Mart{\'\i}n}, {Rusholme}, {Said}, {Salvatelli}, {Salvati}, {Sandri}, {Santos}, {Savelainen}, {Savini}, {Scott}, {Seiffert}, {Serra}, {Shellard}, {Spencer}, {Spinelli}, {Stolyarov}, {Stompor}, {Sudiwala}, {Sunyaev}, {Sutton}, {Suur-Uski}, {Sygnet}, {Tauber}, {Terenzi}, {Toffolatti}, {Tomasi}, {Tristram}, {Trombetti}, {Tucci}, {Tuovinen}, {T{\"u}rler}, {Umana}, {Valenziano}, {Valiviita}, {Van Tent},
  {Vielva}, {Villa}, {Wade}, {Wandelt}, {Wehus}, {White}, {White}, {Wilkinson}, {Yvon}, {Zacchei}, \& {Zonca}}]{2016A&A...594A..13P}
{Planck Collaboration}, {Ade}, P.~A.~R., {Aghanim}, N., {et~al.} 2016, \bibinfo{title}{{Planck 2015 results. XIII. Cosmological parameters},} \aap, 594, A13, \dodoi{10.1051/0004-6361/201525830}

\bibitem[{A. {Rodr{\'\i}guez-Puebla} {et~al.}(2016){Rodr{\'\i}guez-Puebla}, {Behroozi}, {Primack}, {Klypin}, {Lee}, \& {Hellinger}}]{Rodriguez-Puebla2016}
{Rodr{\'\i}guez-Puebla}, A., {Behroozi}, P., {Primack}, J., {et~al.} 2016, \bibinfo{title}{{Halo and subhalo demographics with Planck cosmological parameters: Bolshoi-Planck and MultiDark-Planck simulations},} \mnras, 462, 893, \dodoi{10.1093/mnras/stw1705}

\bibitem[{Y. {Rong} {et~al.}(2025){Rong}, {Wang}, \& {Tang}}]{2025ApJ...983L...3R}
{Rong}, Y., {Wang}, P., \& {Tang}, X.-x. 2025, \bibinfo{title}{{Orthogonal Alignment of Galaxy Group Angular Momentum with Cosmic Filament Spines: An Observational Study},} \apjl, 983, L3, \dodoi{10.3847/2041-8213/adc130}

\bibitem[{A. {Rost} {et~al.}(2020){Rost}, {Stasyszyn}, {Pereyra}, \& {Mart{\'\i}nez}}]{2020MNRAS.493.1936R}
{Rost}, A., {Stasyszyn}, F., {Pereyra}, L., \& {Mart{\'\i}nez}, H.~J. 2020, \bibinfo{title}{{A comparison of cosmological filaments catalogues},} \mnras, 493, 1936, \dodoi{10.1093/mnras/staa320}

\bibitem[{L.~V. {Sales} {et~al.}(2022){Sales}, {Wetzel}, \& {Fattahi}}]{2022NatAs...6..897S}
{Sales}, L.~V., {Wetzel}, A., \& {Fattahi}, A. 2022, \bibinfo{title}{{Baryonic solutions and challenges for cosmological models of dwarf galaxies},} Nature Astronomy, 6, 897, \dodoi{10.1038/s41550-022-01689-w}

\bibitem[{T. {Sousbie}(2011){Sousbie}}]{2011MNRAS.414..350S}
{Sousbie}, T. 2011, \bibinfo{title}{{The persistent cosmic web and its filamentary structure - I. Theory and implementation},} \mnras, 414, 350, \dodoi{10.1111/j.1365-2966.2011.18394.x}

\bibitem[{T. {Sousbie} {et~al.}(2011){Sousbie}, {Pichon}, \& {Kawahara}}]{2011MNRAS.414..384S}
{Sousbie}, T., {Pichon}, C., \& {Kawahara}, H. 2011, \bibinfo{title}{{The persistent cosmic web and its filamentary structure - II. Illustrations},} \mnras, 414, 384, \dodoi{10.1111/j.1365-2966.2011.18395.x}

\bibitem[{V. {Springel}(2010){Springel}}]{2010MNRAS.401..791S}
{Springel}, V. 2010, \bibinfo{title}{{E pur si muove: Galilean-invariant cosmological hydrodynamical simulations on a moving mesh},} \mnras, 401, 791, \dodoi{10.1111/j.1365-2966.2009.15715.x}

\bibitem[{V. {Springel} {et~al.}(2001){Springel}, {White}, {Tormen}, \& {Kauffmann}}]{2001MNRAS.328..726S}
{Springel}, V., {White}, S. D.~M., {Tormen}, G., \& {Kauffmann}, G. 2001, \bibinfo{title}{{Populating a cluster of galaxies - I. Results at z=0},} \mnras, 328, 726, \dodoi{10.1046/j.1365-8711.2001.04912.x}

\bibitem[{V. {Springel} {et~al.}(2008){Springel}, {Wang}, {Vogelsberger}, {Ludlow}, {Jenkins}, {Helmi}, {Navarro}, {Frenk}, \& {White}}]{Springel2008}
{Springel}, V., {Wang}, J., {Vogelsberger}, M., {et~al.} 2008, \bibinfo{title}{{The Aquarius Project: the subhaloes of galactic haloes},} \mnras, 391, 1685, \dodoi{10.1111/j.1365-2966.2008.14066.x}

\bibitem[{V. {Springel} {et~al.}(2018){Springel}, {Pakmor}, {Pillepich}, {Weinberger}, {Nelson}, {Hernquist}, {Vogelsberger}, {Genel}, {Torrey}, {Marinacci}, \& {Naiman}}]{2018MNRAS.475..676S}
{Springel}, V., {Pakmor}, R., {Pillepich}, A., {et~al.} 2018, \bibinfo{title}{{First results from the IllustrisTNG simulations: matter and galaxy clustering},} \mnras, 475, 676, \dodoi{10.1093/mnras/stx3304}

\bibitem[{A. {Storck} {et~al.}(2025){Storck}, {Cadiou}, {Agertz}, \& {Gal{\'a}rraga-Espinosa}}]{2025MNRAS.539..487S}
{Storck}, A., {Cadiou}, C., {Agertz}, O., \& {Gal{\'a}rraga-Espinosa}, D. 2025, \bibinfo{title}{{Exploring the causal effect of cosmic filaments on dark matter haloes},} \mnras, 539, 487, \dodoi{10.1093/mnras/staf523}

\bibitem[{E. {Tempel} {et~al.}(2014){Tempel}, {Stoica}, {Mart{\'\i}nez}, {Liivam{\"a}gi}, {Castellan}, \& {Saar}}]{2014MNRAS.438.3465T}
{Tempel}, E., {Stoica}, R.~S., {Mart{\'\i}nez}, V.~J., {et~al.} 2014, \bibinfo{title}{{Detecting filamentary pattern in the cosmic web: a catalogue of filaments for the SDSS},} \mnras, 438, 3465, \dodoi{10.1093/mnras/stt2454}

\bibitem[{P. {Virtanen} {et~al.}(2020){Virtanen}, {Gommers}, {Oliphant}, {Haberland}, {Reddy}, {Cournapeau}, {Burovski}, {Peterson}, {Weckesser}, {Bright}, {van der Walt}, {Brett}, {Wilson}, {Millman}, {Mayorov}, {Nelson}, {Jones}, {Kern}, {Larson}, {Carey}, {Polat}, {Feng}, {Moore}, {VanderPlas}, {Laxalde}, {Perktold}, {Cimrman}, {Henriksen}, {Quintero}, {Harris}, {Archibald}, {Ribeiro}, {Pedregosa}, {van Mulbregt}, \& {SciPy 1. 0 Contributors}}]{Virtanen2020}
{Virtanen}, P., {Gommers}, R., {Oliphant}, T.~E., {et~al.} 2020, \bibinfo{title}{{SciPy 1.0: fundamental algorithms for scientific computing in Python},} Nature Methods, 17, 261, \dodoi{10.1038/s41592-019-0686-2}

\bibitem[{P. {Wang} {et~al.}(2018){Wang}, {Guo}, {Kang}, \& {Libeskind}}]{Wang2018}
{Wang}, P., {Guo}, Q., {Kang}, X., \& {Libeskind}, N.~I. 2018, \bibinfo{title}{{The Spin Alignment of Galaxies with the Large-scale Tidal Field in Hydrodynamic Simulations},} \apj, 866, 138, \dodoi{10.3847/1538-4357/aae20f}

\bibitem[{W. {Wang} {et~al.}(2025){Wang}, {Wang}, {Rong}, {Wang}, \& {Tang}}]{2025JCAP...10..095W}
{Wang}, W., {Wang}, P., {Rong}, Y., {Wang}, H.-d., \& {Tang}, X.-x. 2025, \bibinfo{title}{{Galaxy group spin alignment with cosmic filament in the TNG simulation},} \jcap, 2025, 095, \dodoi{10.1088/1475-7516/2025/10/095}

\bibitem[{W. {Wang} {et~al.}(2024){Wang}, {Wang}, {Guo}, {Kang}, {Libeskind}, {Gal{\'a}rraga-Espinosa}, {Springel}, {Kannan}, {Hernquist}, {Pakmor}, {Yu}, {Bose}, {Guo}, {Yu}, \& {Hern{\'a}ndez-Aguayo}}]{2024MNRAS.532.4604W}
{Wang}, W., {Wang}, P., {Guo}, H., {et~al.} 2024, \bibinfo{title}{{The boundary of cosmic filaments},} \mnras, 532, 4604, \dodoi{10.1093/mnras/stae1801}

\bibitem[{W. {Xu} {et~al.}(2020){Xu}, {Guo}, {Zheng}, {Gao}, {Lacey}, {Gu}, {Liao}, {Shao}, {Mao}, {Zhang}, \& {Chen}}]{2020MNRAS.498.1839X}
{Xu}, W., {Guo}, Q., {Zheng}, H., {et~al.} 2020, \bibinfo{title}{{Galaxy properties in the cosmic web of EAGLE simulation},} \mnras, 498, 1839, \dodoi{10.1093/mnras/staa2497}

\bibitem[{Q.-R. {Yang} {et~al.}(2025){Yang}, {Zhu}, {Yu}, {Mo}, {Zheng}, \& {Feng}}]{2025ApJ...989..187Y}
{Yang}, Q.-R., {Zhu}, W., {Yu}, G.-Y., {et~al.} 2025, \bibinfo{title}{{On the Width and Profiles of Cosmic Filaments},} \apj, 989, 187, \dodoi{10.3847/1538-4357/adeca3}

\bibitem[{G. {Yu} {et~al.}(2025){Yu}, {Zhu}, {Yang}, {Mo}, {Luan}, \& {Feng}}]{2025ApJ...986..193Y}
{Yu}, G., {Zhu}, W., {Yang}, Q.-R., {et~al.} 2025, \bibinfo{title}{{Impact of the Cosmic Web on the Properties of Galaxies in IllustrisTNG Simulations},} \apj, 986, 193, \dodoi{10.3847/1538-4357/adc80f}

\bibitem[{D. {Zakharova} {et~al.}(2023){Zakharova}, {Vulcani}, {De Lucia}, {Xie}, {Hirschmann}, \& {Fontanot}}]{2023MNRAS.525.4079Z}
{Zakharova}, D., {Vulcani}, B., {De Lucia}, G., {et~al.} 2023, \bibinfo{title}{{The filament determination depends on the tracer: comparing filaments based on dark matter particles and galaxies in the GAEA semi-analytical model},} \mnras, 525, 4079, \dodoi{10.1093/mnras/stad2562}

\bibitem[{D. {Zakharova} {et~al.}(2024){Zakharova}, {Vulcani}, {De Lucia}, {Finn}, {Rudnick}, {Combes}, {Castignani}, {Fontanot}, {Jablonka}, {Xie}, \& {Hirschmann}}]{2024A&A...690A.300Z}
{Zakharova}, D., {Vulcani}, B., {De Lucia}, G., {et~al.} 2024, \bibinfo{title}{{Virgo Filaments: III. The gas content of galaxies in filaments as predicted by the GAEA semi-analytic model},} \aap, 690, A300, \dodoi{10.1051/0004-6361/202450825}

\bibitem[{J. {Zavala} \& C.~S. {Frenk}(2019){Zavala} \& {Frenk}}]{Zavala2019}
{Zavala}, J., \& {Frenk}, C.~S. 2019, \bibinfo{title}{{Dark Matter Haloes and Subhaloes},} Galaxies, 7, 81, \dodoi{10.3390/galaxies7040081}

\bibitem[{Y. {Zhang} {et~al.}(2025){Zhang}, {Yang}, {Guo}, {Wang}, \& {Shi}}]{2025MNRAS.539.1692Z}
{Zhang}, Y., {Yang}, X., {Guo}, H., {Wang}, P., \& {Shi}, F. 2025, \bibinfo{title}{{Galaxy and halo properties around cosmic filaments from Sloan Digital Sky Survey Data Release 7 and the ELUCID simulation},} \mnras, 539, 1692, \dodoi{10.1093/mnras/staf611}

\bibitem[{H. {Zheng} {et~al.}(2022){Zheng}, {Liao}, {Hu}, {Gao}, {Grand}, {Gu}, \& {Guo}}]{2022MNRAS.514.2488Z}
{Zheng}, H., {Liao}, S., {Hu}, J., {et~al.} 2022, \bibinfo{title}{{The impact of filaments on dwarf galaxy properties in the Auriga simulations},} \mnras, 514, 2488, \dodoi{10.1093/mnras/stac1476}

\bibitem[{W. {Zhu} {et~al.}(2022){Zhu}, {Zhang}, \& {Feng}}]{2022ApJ...924..132Z}
{Zhu}, W., {Zhang}, F., \& {Feng}, L.-L. 2022, \bibinfo{title}{{Impact of Cosmic Filaments on the Gas Accretion Rate of Dark Matter Halos},} \apj, 924, 132, \dodoi{10.3847/1538-4357/ac37b9}

\end{thebibliography}
\bibliographystyle{aasjournalv7}

\end{document}